%
%
\magnification=\magstep1
\baselineskip=11pt plus .1pt minus .1pt
\hsize=12.5truecm
\vsize=19.0truecm  
\hfuzz=5pt\vfuzz=5pt
\tolerance=1000
\overfullrule=0pt
\parskip=0pt
\abovedisplayskip=3 mm plus6pt minus 4pt
\belowdisplayskip=3 mm plus6pt minus 4pt
\abovedisplayshortskip=0mm plus6pt minus 2pt
\belowdisplayshortskip=2 mm plus4pt minus 4pt
\predisplaypenalty=0
\clubpenalty=10000
\widowpenalty=10000
\parindent=2em
%
%
\font\pgnumfont=cmr9
\font\headlinefont=cmti9
 \font\titlefont=cmbx10
\font\authorfont=cmr10
\font\addressfont=cmti9
\font\datefont=cmr9
\font\sumfont=cmr9

\font\absfont=cmbx9
\font\secfont=cmr10
\font\subsecfont=cmti10
\font\subsubsecfont=cmr10
\font\figfont=cmr9
\font\figheadfont=cmbx9
\font\tabfont=cmr9
\font\tabheadfont=cmbx9
\font\mainfont=cmr10
\font\petitrm=cmr9

%
%
%
\newtoks\TITLE \newtoks\AUTHOR \newtoks\ADDRESS \newtoks\SUMMARY
\newdimen\sumindent \sumindent=\parindent
\newtoks\KEYWORDS \newtoks\SUBMITTED \newtoks\ACCEPTED
\newtoks\SENDOFF
%

%
%
\newtoks\firstpage
\let\firstpage=Y
\newtoks\AUTHORHEAD \newtoks\ARTHEAD \newtoks\VOLUME \newtoks\PAGES
\if!\the\AUTHORHEAD!\AUTHORHEAD={\the\AUTHOR}\fi
\if!\the\ARTHEAD!\ARTHEAD={\the\TITLE}\fi
\footline={\hfil}
\headline={\ifodd\pageno\rightheadline \else\leftheadline\fi}
\def\leftheadline{\if Y\firstpage\firsthead\global\let\firstpage=N
  \else\lefthead\fi}
\def\rightheadline{\if Y\firstpage\firsthead\global\let\firstpage=N
  \else\righthead\fi}
\def\lefthead{\pgnumfont\number\pageno\hfil\headlinefont\the\AUTHORHEAD}
\def\righthead{\headlinefont\the\ARTHEAD\hfil\pgnumfont\number\pageno}
\def\firsthead{\headlinefont Baltic Astronomy, vol.~\the\VOLUME,
\the\PAGES,~\the\year .\hfil}
\voffset=2\baselineskip 
%

\newdimen\oldbaselineskip \oldbaselineskip=\baselineskip
\def\test#1{\newlinechar=`@\if!\the#1! \message{#1 not given@}\fi}%
\def\printheader{
  \parindent=0pt
  \null\vskip1.cm
  \test{\TITLE}
  \vbox{\baselineskip=15pt
    \titlefont\the\TITLE
    }
  \vskip8mm plus8mm
  \test{\AUTHOR}
  \authorfont\the\AUTHOR
  \vskip2mm
  \test{\ADDRESS}
  \addressfont\the\ADDRESS
  \vskip2mm
  \test{\SUBMITTED}
  \line{\datefont Received \the\SUBMITTED
    \if!\the\ACCEPTED!\else, accepted \the\ACCEPTED\fi\hfill}
  \vskip4mm plus4mm
{
\leftskip=\sumindent\parindent=0pt
    \parskip=5pt
    \absfont Abstract.
    \test{\SUMMARY}
    \sumfont\the\SUMMARY\par
    \absfont Key words:
    \test{\KEYWORDS}
    \sumfont\the\KEYWORDS\par
    }
  \sumfont
  \if!\the\SENDOFF!\else\footnote{}{Send offprint requests to:
 \the\SENDOFF}\fi
  \parindent=2em
  }
%
%
\newdimen\uppergap \newdimen\lowergap
\uppergap=5mm \lowergap=3mm
\newdimen\secind \newdimen\subsecind \newdimen\subsubsecind
\setbox0=\hbox{\secfont 9. }\secind=\wd0
\setbox0=\hbox{\subsecfont 9.9. }\subsecind=\wd0
\setbox0=\hbox{\subsubsecfont 9.9.9. }\subsubsecind=\wd0
\def\section#1{\goodbreak\par\vskip\uppergap
  \noindent\hangindent\secind\hangafter=1\secfont#1
  \vskip\lowergap\mainfont\par\nobreak}
\def\subsection#1{\goodbreak\par\vskip\uppergap
  \noindent\hangindent\subsecind\hangafter=1\subsecfont#1
  \vskip\lowergap\mainfont\par\nobreak}
\def\subsubsection#1{\goodbreak\par\vskip\uppergap
  \noindent\hangindent\subsubsecind\hangafter=1\subsubsecfont#1
  \vskip\lowergap\mainfont\par\nobreak}
\def\altaffilmark#1{{$^{#1}$}}
\def\altaffiltext#1#2{\moveright4mm\vbox{\hsize100mm{\item{$^{#1}$}{#2}}}}
\def~{\penalty10000\thinspace}
%
%
\def\WFigure#1#2#3{\goodbreak\midinsert\vbox{
  \null\centerline{#2}\vskip1.5truemm
  \figheadfont\indent Fig.~#1.\figfont\ #3
  \par\mainfont
  }\endinsert}
\def\bcaption#1#2{\figheadfont\indent Fig.\ts #1.\figfont\ #2\par\mainfont}
%
\newdimen\tabind
\setbox0=\hbox{\tabheadfont Table 55.} \tabind=\wd0

%
%
\def\References{\vskip\uppergap
\line{\secfont REFERENCES\hfill}
  \vskip0.8\lowergap
 \petitrm
  }
\def\ref{\goodbreak
\hangindent12pt\hangafter=1
\noindent\ignorespaces}
\def\endref{\egroup}
%
%
\def\byebye{\egroup\par\vfill\supereject\end}
%
%

%
%

\def\utw{\smash{\rlap{\lower5pt\hbox{$\sim$}}}}
\def\udtw{\smash{\rlap{\lower6pt\hbox{$\approx$}}}}

\def\fm{\hbox{$.\!\!^{\rm m}$}}

\font\tbold=cmbx9
\font\tabfont=cmr9
\def\tablerule{\noalign{\vskip.9ex}\noalign{\hrule}\noalign{\vskip.7ex}}
\def\huad{\vrule height0pt depth0pt width5pt}  


\def\ddown{\lower2.5ex\hbox}
\def\ddow{\lower1.7ex\hbox}
\def\down{\lower1ex\hbox}
\def\uppp{\raise1ex\hbox}
\def\dnnn{\lower1ex\hbox}
\def\uuppp{\raise2ex\hbox}

\def\ts{\thinspace}
\def\(o-c){$O-C$}


\def\angstr{A\kern-.56em\raise1.9ex\hbox{$\scriptscriptstyle\circ$}$\,$}

\newdimen\free\newdimen\shift
\def\Entry#1#2#3{\par\goodbreak\smallskip%
  \setbox1=\vbox{\advance\hsize by-10mm\parindent=0pt
    \def\\{\par}%
    \it#1. \rm#2}
  \line{\box1\hfill#3}\smallskip
}%
\newdimen\savesize

\def\shiftfigure #1#2#3#4#5{
    \vbox to #2 { \ifodd #5 \rightskip#4 \else\leftskip#4 \fi
                  \null\vfil
                  \figheadfont Fig.~#1.\figfont #3
                  \medskip
                }
                          }

\year2000

\input psfig.sty
\TITLE={GALAXY SURFACE PHOTOMETRY}
\VOLUME={8}
\PAGES={535--574}              
\pageno=535                      
\year=1999
\AUTHORHEAD={B. Milvang-Jensen \& I. J{\o}rgensen}

\AUTHOR{
Bo Milvang-Jensen\altaffilmark{1}$^{\rm ,}$\altaffilmark{2}
and
Inger J{\o}rgensen\altaffilmark{3}
}

\ARTHEAD={Galaxy surface photometry} 

\ADDRESS={%
\altaffiltext{1}{%
Copenhagen University Observatory,
2100 Copenhagen {\O},
Denmark,
{\tt milvang@astro.ku.dk}
}
\altaffiltext{2}{%
School of Physics and Astronomy,
University of Nottingham,
University Park,
NG7 2RD Nottingham,
UK
(postal address for BMJ)
}
\altaffiltext{3}{%
Gemini Observatory,
670 N. Aohoku Pl.,
Hilo, Hawaii 96720,
USA,
{\tt ijorgensen@gemini.edu}
}
}

\SUMMARY={%
We describe galaxy surface photometry based on fitting ellipses
to the isophotes of the galaxies.
Example galaxies with different isophotal shapes are used to illustrate
the process, including how the deviations from elliptical isophotes
are quantified using Fourier expansions.
We show how the definitions of the Fourier coefficients
employed by different authors are linked.
As examples of applications of surface photometry
we discuss the determination of the relative disk luminosities
and the inclinations for E and S0 galaxies.
We also describe the color-magnitude and color-color relations.
When using both near-infrared and optical photometry,
the age--metallicity degeneracy may be broken.
Finally we discuss the Fundamental Plane where surface photometry is
combined with spectroscopy.
It is shown how the FP can be used as a sensitive tool
to study galaxy evolution.
}

\KEYWORDS={%
%
techniques: photometric --
galaxies: photometry --
galaxies: fundamental parameters --
galaxies: elliptical and lenticular, cD --
galaxies: stellar content --
galaxies: evolution
}
\SUBMITTED={March 3, 2000}
\printheader


\def\fm{\hbox{$.\!\!^{\rm m}$}}
\def\MBT{{M_{\rm B_T}}}
\def\marc{{^{\rm m} /{\rm arcsec}^{2}}}
\def\cfour{{c_{\rm 4}}}
\def\csix{{c_{\rm 6}}}
\def\mT{{m_{\rm T}}}
\def\epsiso{{\epsilon_{21.85}}}
\def\LD{{L_{\rm D}}}
\def\Ltot{{L_{\rm tot}}}
\def\aeB{{a_{\rm eB}}}
\def\aeD{{a_{\rm eD}}}
\def\Ho50{{H_{\rm 0}=50~{\rm km\, s^{-1}\, Mpc^{-1}} }}

\def\barcfour{\langle c_4 \rangle}
\def\barcsix{\langle c_6 \rangle}
\def\Ie{\langle I \rangle_{\rm e}}
\def\mue{\langle \mu \rangle_{\rm e}}
\def\barsn{\langle s_n \rangle}
\def\barcn{\langle c_n \rangle}

\def\re{r_{\rm e}}
\def\dInorm{\Delta I_{\rm norm}}
\def\MrT{M_{\rm r_T}}
\def\rmin{{r_{\rm min}}}
\def\rmax{{r_{\rm max}}}
\def\Rangle{\langle R \rangle}
\def\Vanglesq{\langle V^2 \rangle}
\def\kR{k_{\rm R}}
\def\kV{k_{\rm V}}
\def\kL{k_{\rm L}}
\def\kS{k_{\rm S}}


\dimen0=\hsize
\divide\dimen0 by 1000

\dimen1=\dimen0
\dimen2=\dimen0
\dimen3=\dimen0
\dimen6=\dimen0
\dimen8=\dimen0
\dimen9=\dimen0
\multiply\dimen6 by 468 
\multiply\dimen1 by 490 
\multiply\dimen3 by 784 %
\multiply\dimen8 by 936 
\multiply\dimen9 by 980 
\multiply\dimen2 by 999 


\section{1. INTRODUCTION}

Surface photometry of galaxies is a technique to quantitatively
describe the light distribution of the galaxies,
as recorded in 2-dimensional images.
This paper focuses on the techniques used to derive the surface
photometry and presents some examples of scientific applications.
The paper is a summary of the lectures given at the summer school,
and is not intended as a complete review of the topic.
Surface photometry of galaxies has previously been reviewed by
Kormendy \& Djorgovski (1989) and Okamura (1988).

The techniques and the software used for performing surface photometry 
of galaxies are described in Section 2.
Surface photometry has many applications.
In Section 3 we give an example of such an application,
namely the determination of disk luminosities and inclinations of
E and S0 galaxies.
This determination is based on surface photometry --
ellipticities and deviations from elliptical shape -- alone.

 From surface photometry in several passbands we can derive the
colors of the galaxies.
These colors provide information about the ages and the metal content of
the stellar populations in the galaxies.
In order to study this, stellar population models are needed. These
models are the subject of Section 4.
In Section 5, we then present examples of the color-magnitude
and color-color relations of galaxies.

Although the topic of this summer school is photometry, we will 
nevertheless show an example of the science that can be carried out
when surface photometry is combined with spectroscopy. 
The chosen example is the relation known as the 
Fundamental Plane for E and S0 galaxies, which we discuss in Section 6.

We end this description of surface photometry with a few
suggestions for future projects that can be carried
out based on surface photometry only, see Section 7.

Throughout the paper we use
$H_0 = 50\,{\rm km}\,{\rm s}^{-1}\,{\rm Mpc}^{-1}$.

\vskip2truemm

\section{2. SURFACE PHOTOMETRY}

Surface photometry is used to study {\it extended objects\/},
such as galaxies, as opposed to {\it point sources}, such as stars.
 From an image of a point source, only its total magnitude can be derived.
 From an image of a galaxy, it is possible to determine a number of quantities.
Some of these quantities are derived from surface photometry,
such as how the intensity and ellipticity vary with radius.
Other quantities are determined by other methods -- the morphological type,
for example, is determined from visual inspection of the image.
There also exist schemes to do automated morphological classification,
e.g.\ Abraham et al.\ (1994, 1996); Naim et al.\ (1995).

In this context, we are talking about galaxies where the
individual stars cannot be resolved in the images.
This is for example the case for the HydraI (Abell 1060) cluster,
which is a nearby cluster at a distance of $\sim 80$ Mpc.
At that distance, even the Hubble Space Telescope (HST) cannot resolve
individual stars in the galaxies.

Photometry of crowded stellar fields, such as globular 
clusters, is not considered surface photometry.
Surface photometry is used when the magnitudes of the individual stars 
(typically in a galaxy) cannot be measured,
but only their smooth integrated light.

\subsection{2.1 Ellipse fitting}

Surface photometry of galaxies is usually done by fitting ellipses to the
isophotes. This choice is motivated by that fact that the isophotes
of galaxies are not far from ellipses. This is especially the case for
elliptical (E) and lenticular (S0) galaxies.
In this paper we will concentrate on E and S0 galaxies.
We will also limit our discussion to data obtained with CCDs 
(Charge-Coupled Devices).

There exist several software packages for deriving surface photometry.
For the exercise related to the lectures given at the summer school
we have chosen the ELLIPSE task in the ISOPHOTE package
(see Busko 1996).
This task is based on the ellipse fitting algorithm used in
the GASP package by Cawson (1983; see also Davis et al.\ 1985),
the code for which was later rewritten by Jedrzejewski (1987).
ISOPHOTE is a part of the external IRAF\footnote{$^\ast$}
{IRAF is distributed by the National Optical Astronomy Observatories,
which are operated by the Association of Universities for Research
in Astronomy, Inc., under cooperative agreement with the National
Science Foundation.
See also {\tt http://iraf.noao.edu/}}
package STSDAS\footnote{$^{\ast\ast}$}
{STSDAS is distributed by the Space Telescope Science Institute,
operated by AURA, Inc., under NASA contract NAS 5-26555.
See also {\tt http://ra.stsci.edu/STSDAS.html}}.
We chose this software for the exercise since it is publicly available, and 
because it includes some documentation.
The examples of surface photometry presented in this section are also
based on the ISOPHOTE package.
Our aim is to describe the basic principles of
surface photometry based on ellipse fitting, and the choice of software
package is not critical for that purpose.
A comparison of the results obtained with ELLIPSE
with those obtained with other software packages 
is beyond the scope of this paper.

To illustrate how surface photometry on E and S0 galaxies is derived we have chosen 
three example galaxies, see Figure~1 and Table~1.
We will refer to these galaxies as the `pure E', the `S0' and the `boxy E',
respectively.
The name `pure E' refers to the fact that the isophotes of this galaxy are
almost perfect ellipses.
For the `boxy E', the isophotes are box shaped.
The pure E has been morphologically classified as E3/S0.
The boxy E has been morphologically classified as SB(rs)0(0)
(barred S0 with rings).
For our purpose of illustrating surface photometry these morphological
types are not important.

\WFigure{1}{%
\psfig{file=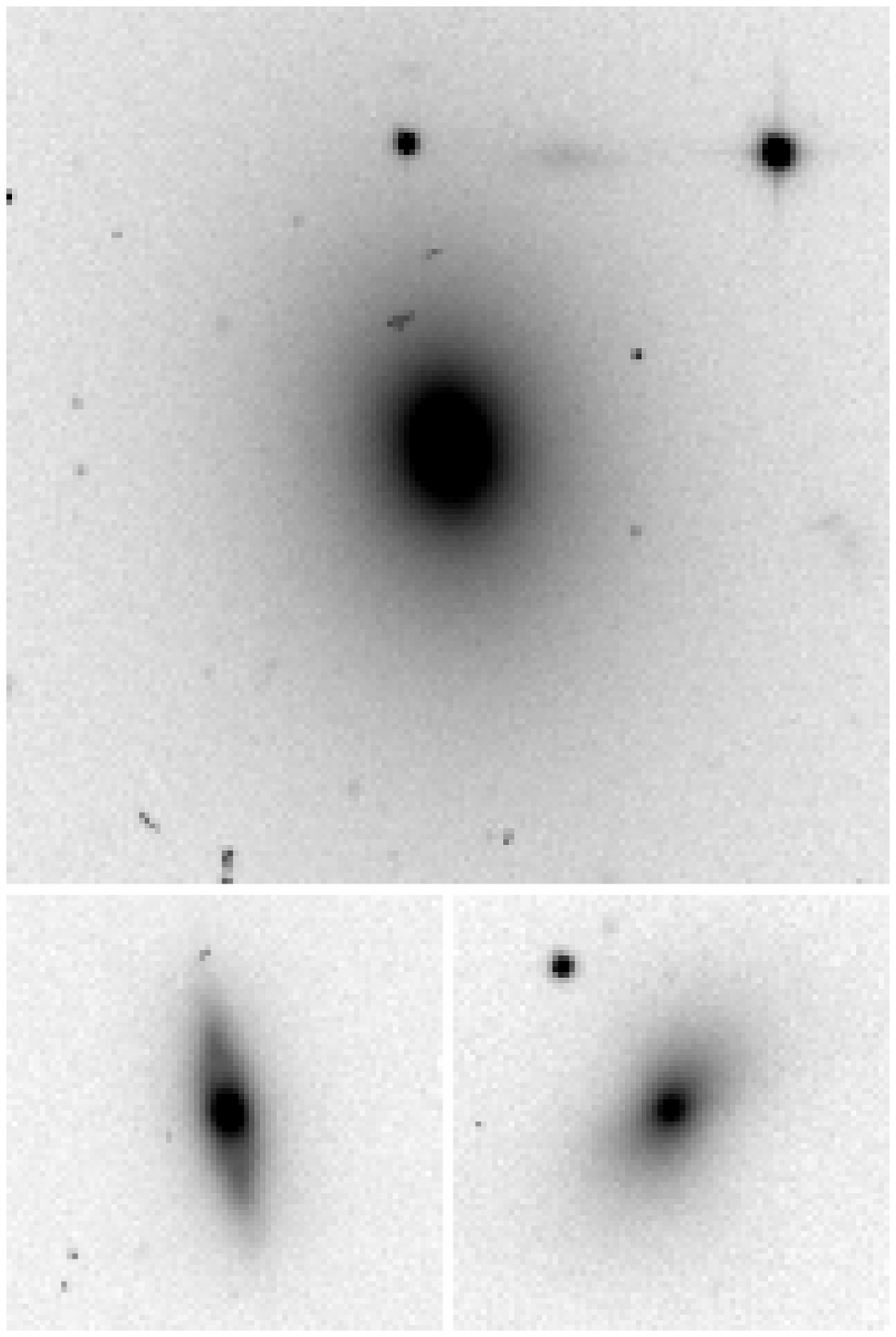,width=\dimen8,clip=}
}
{The three example galaxies -- pure E (top), S0 (left), boxy E (right).
The width of the montage is 32 kpc. North is down and east is to the right.}

\midinsert
$$\vbox{\tabfont
\halign{
\huad #\hfil&           
\huad #\hfil&           
\huad #\hfil&           
\huad \hfil#&           
\huad \hfil#\cr         
{\tbold Table 1.}
&\multispan{4}{\ The three example galaxies\hfil}\cr
\tablerule
Description & Name           & Morph.\ type & $\MrT$       & $\re$   \cr
\tablerule
Pure E      & R347 / IC2597  & E3/S0        & $-23.35$ mag & 8.9 kpc \cr
S0          & R338           & S0(5)        & $-20.74$ mag & 1.9 kpc \cr
Boxy E      & R245           & SB(rs)0(0)   & $-20.94$ mag & 4.8 kpc \cr
\tablerule
\multispan{5}{\vbox{{\bf Note}:\
Name and morphological type are from Richter (1989).
Total absolute Gunn $r$ magnitude ($\MrT$) and effective radius ($\re$)
are from Milvang-Jensen \& J{\o}rgensen (2000),
based on $H_0 = 50\,{\rm km}\,{\rm s}^{-1}\,{\rm Mpc}^{-1}$.}}\cr
}}
$$
\endinsert

\WFigure{2}{%
\psfig{file=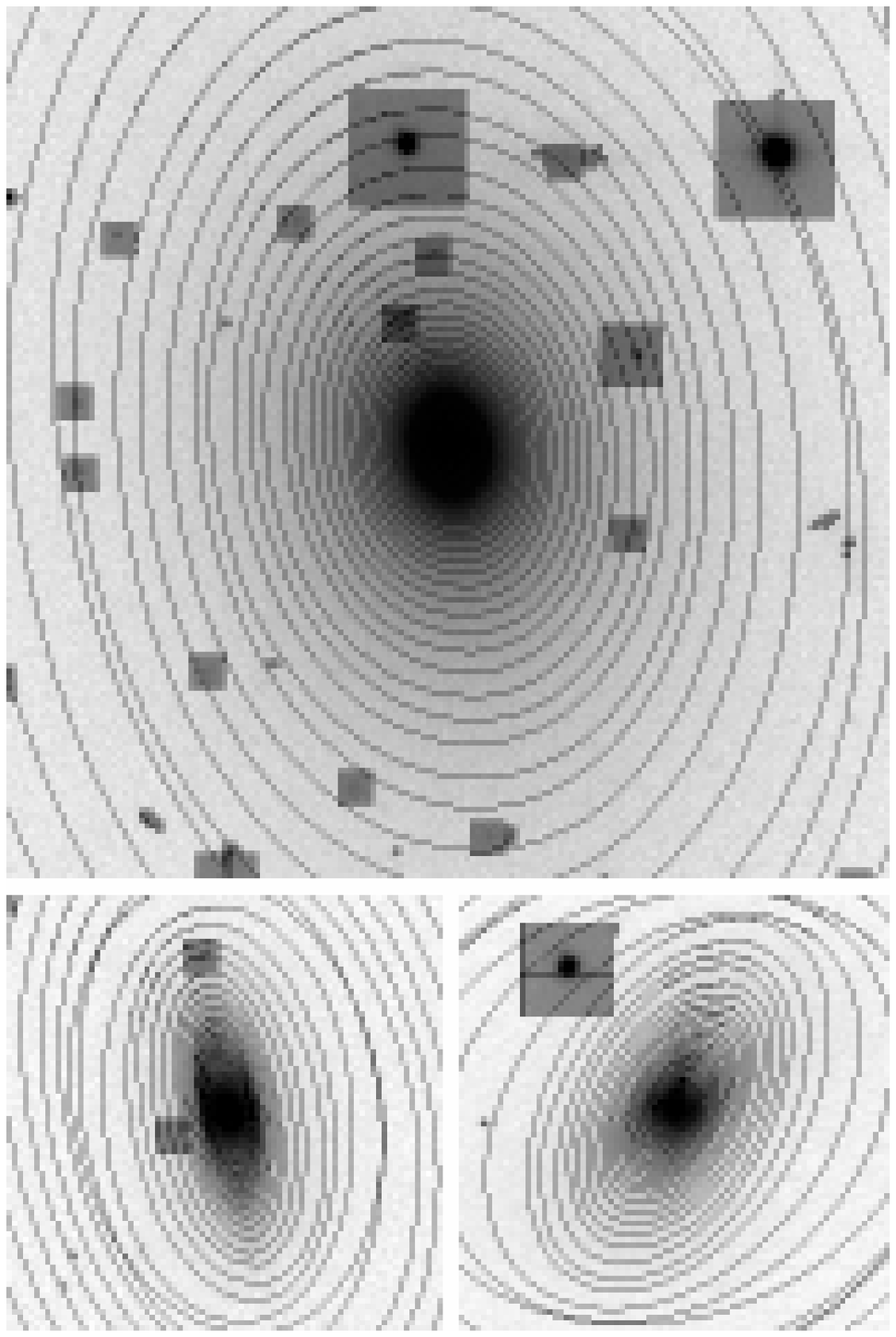,width=\dimen8,clip=}
}
{The three example galaxies, with best-fitting ellipses,
and with pixels contaminated by other objects flagged (the hatched 
areas).}

\WFigure{3}{%
\psfig{file=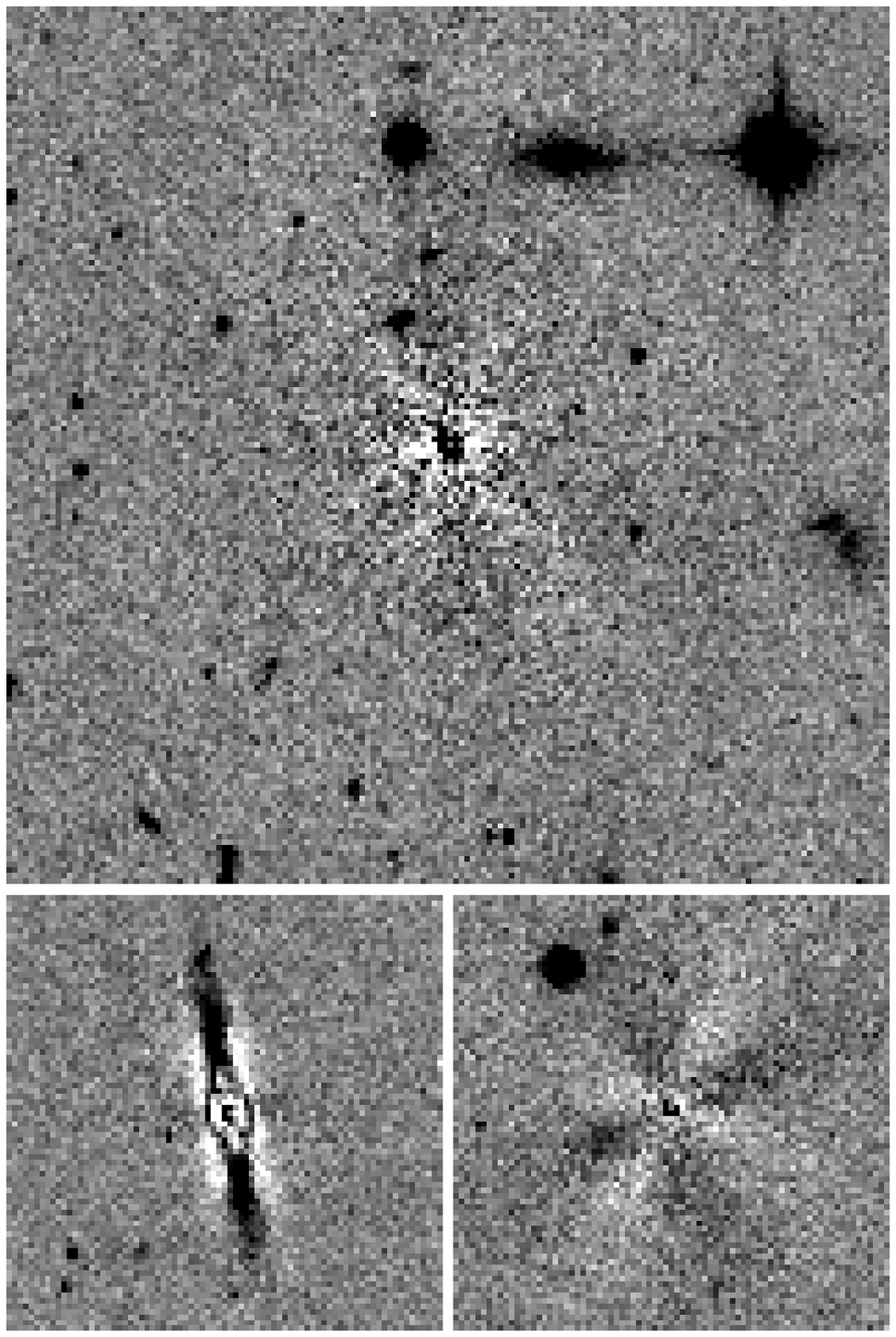,width=\dimen8,clip=}
}
{The three example galaxies:
residual images (i.e.\ the original image minus the model image).
The intensity scaling is linear, and the cuts are symmetrical around zero.
Black represents positive residuals.
}

The example galaxies are all members of the HydraI (Abell 1060) cluster.
The images used are 300 sec Gunn $r$ exposures 
($\lambda_{\rm eff} = 6550 \, {\rm \AA}$, Thuan \& Gunn 1976)
obtained with the Danish 1.5-meter Telescope, La~Silla, Chile.
The spatial scales, as well as the grey-scales
of the images of the three galaxies in the figures are identical,
allowing a direct visual comparison between the three galaxies.
The intensity scaling in the images shown on Figure~1 is logarithmic.
We have given lengths in kpc rather than arcsec.
To calculate lengths in arcsec, use
  $\log(\ell/{\rm arcsec}) = \log(\ell/{\rm kpc}) + 0.407$.
%
The corresponding distance modulus for the cluster is
$(m-M) = 34.60$ mag.

In doing the surface photometry, the first step is to identify the
other objects in the image (other galaxies, stars, and cosmic ray events)
and mask (flag) these. This is shown in Figure~2.
The masked pixels are not used in the surface photometry of the galaxy.
The masking shown in Figure~2 uses squares for the masking of objects.
A masking using circles is of course also possible, and that would
mask fewer `uncontaminated' pixels.

Once the masking is done, we need only provide the surface photometry 
task the approximate location of the center of the galaxy.
The task then fits ellipses to the intensities in the image.
This is done at a number of discrete radii,
as also shown in Figure~2.
For the ELLIPSE task, these discrete radii are specified by the rule
that the different semi-major axis lengths $a$ are spaced by a factor of $1.1$.

The center ($x$,$y$) and the shape (ellipticity $\epsilon$ and position angle)
of the ellipses are kept as free parameters in the fit
for semi-major axes at which the signal-to-noise ratio is sufficiently
high. For larger semi-major axes, where the signal-to-noise ratio
is lower, the center and shape of the ellipses are 
fixed.

\topinsert
\vbox{%
\centerline{%
\psfig{file=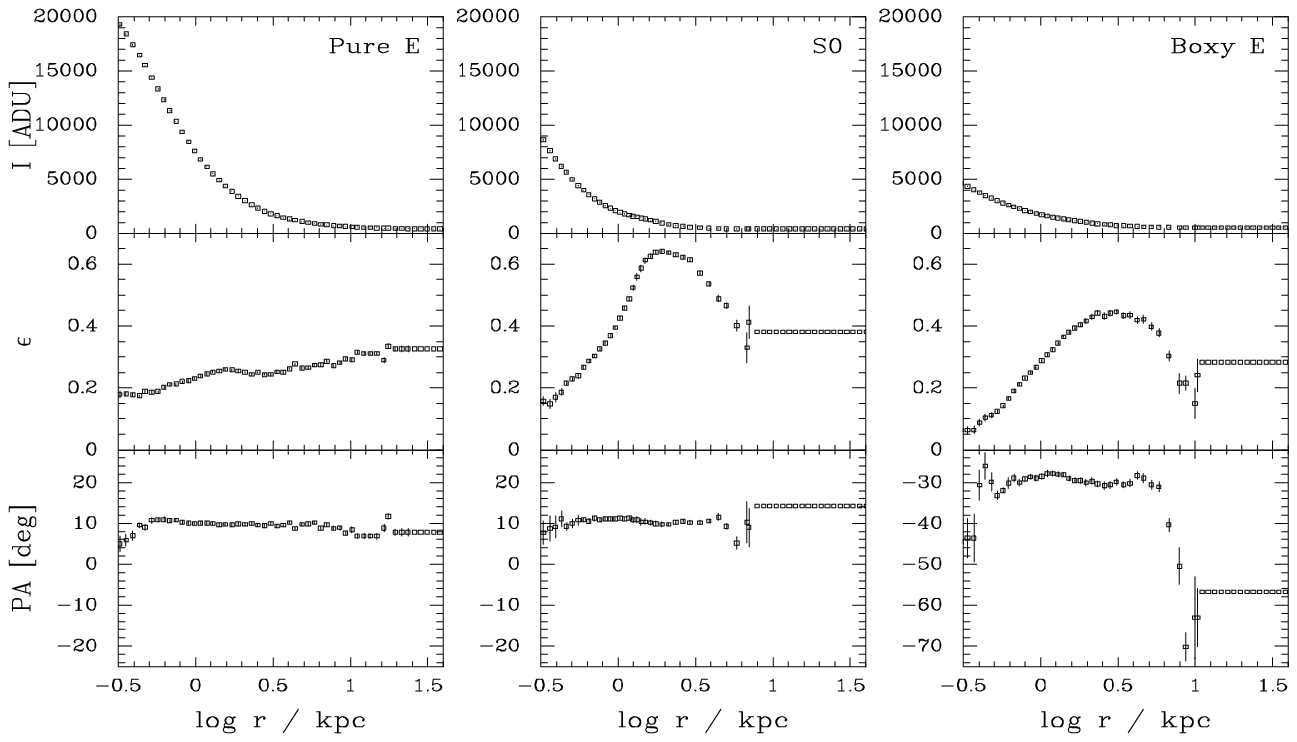,width=\dimen2,clip=}
}
\bcaption{4}
{Radial profiles of intensity, ellipticity and position angle
for the three example galaxies.
The plotted range in equivalent radii $r$ is 0.3--40 kpc (0.8$''$--102$''$).}
}
\endinsert


We get two types of output from the ellipse fit.
One type is the residual image,
which is the difference between the original image and the model image
based on the best-fitting ellipses.
The residual images for the three example galaxies are shown in Figure~3.

For the pure E, the residuals are fairly small.
For the S0 in particular, and for the boxy E, the residuals are larger.
The pure E is much brighter than the other two galaxies,
so in relative terms the residuals for the pure E are much smaller
than for the other two galaxies.
For the pure E, very little structure is seen in the residual image.
The residual images of the S0 and the boxy E show clear structures.
In Section 2.2 we will discuss how to quantify these structures.

The other type of output from the ellipse fit is the {\it radial profiles\/}
of a number of quantities.
I.e., for each ellipse we get the intensity, center,
ellipticity, position angle,
and measures of the deviations from perfect elliptical isophotes
(see Section 2.2), 
as well as the uncertainties for all these quantities.
In Figure~4 we show the radial profiles of
intensity, ellipticity and position angle
for the three example galaxies.
The position angles shown in Figure~4 are measured counter-clockwise from
the $y$-axis of the images. The ELLIPSE task adopts this definition
of position angles.
The standard (astronomical) definition of position angles is from 
north through east.
Our images have north down and east to the right,
and thus need to be rotated by 180$^\circ$ to have the $y$-axis
poiting towards north.
However, position angles are only unique to within 180$^\circ$ since
the major axis of a galaxy does not have a direction (as opposed to a
coordinate axis).
Therefore, the position angles shown in Figure~4 {\it are\/} expressed in the
standard way.
It also follows, for example, that the position angle of the inner isophotes
of the boxy galaxy could be said to be 150$^\circ$ as well as $-30^\circ$.

In Figure~4, we have plotted the different quantities against
the logarithm of the {\it equivalent radius\/} $r$.
The equivalent radius is defined by $r = \sqrt{ab}$,
where $a$ and $b$ are the semi-major and semi-minor axis of the ellipse,
respectively.
A circle with radius $r$ has an area equivalent to an ellipse with
axes $a$ and $b$, hence the name {\it equivalent\/} radius.
The program used for the illustrations in this section, ELLIPSE,
uses $a$ to characterize the size of the ellipses.
However, since the ellipticity $\epsilon$ of the ellipse is defined as
$\epsilon = 1 - b/a$, it follows that
$r$ can be calculated from $a$ and $\epsilon$ as
$r = a \sqrt{1-\epsilon}$.

 From Figure~4 it is seen how the ellipticity and position angle are free
parameters until a certain radius were their values are fixed.
The ellipticity is seen to vary with radius for all three galaxies.
The position angle for the boxy galaxy is seen to vary rapidly in the outer parts,
which is also seen in Figure~2. This behavior is known as an isophote twist.

\subsection{2.2 Quantifying the deviations from elliptical shapes}

As we saw from the residual images in Figure~3,
the isophotes of E and S0 galaxies are not always perfectly elliptical.
We wish to quantify these deviations from elliptical shapes.
To illustrate how this is done,
we have chosen an {\it example ellipse\/} for each of
the three example galaxies.
The three example ellipses are shown in Figure~5, 6 and 7 --
they are overlayed both on the original and on the residual images of the
example galaxies.

In Figure~5, 6 and 7, we also plot the quantity
$$
\dInorm \equiv
{I - I_0 \over {r \cdot {\bigl|{{\rm d}I \over {\rm d}r}\bigr|}}}
\enspace . \eqno(1)
$$
\midinsert
\vbox{%
\centerline{%
\psfig{file=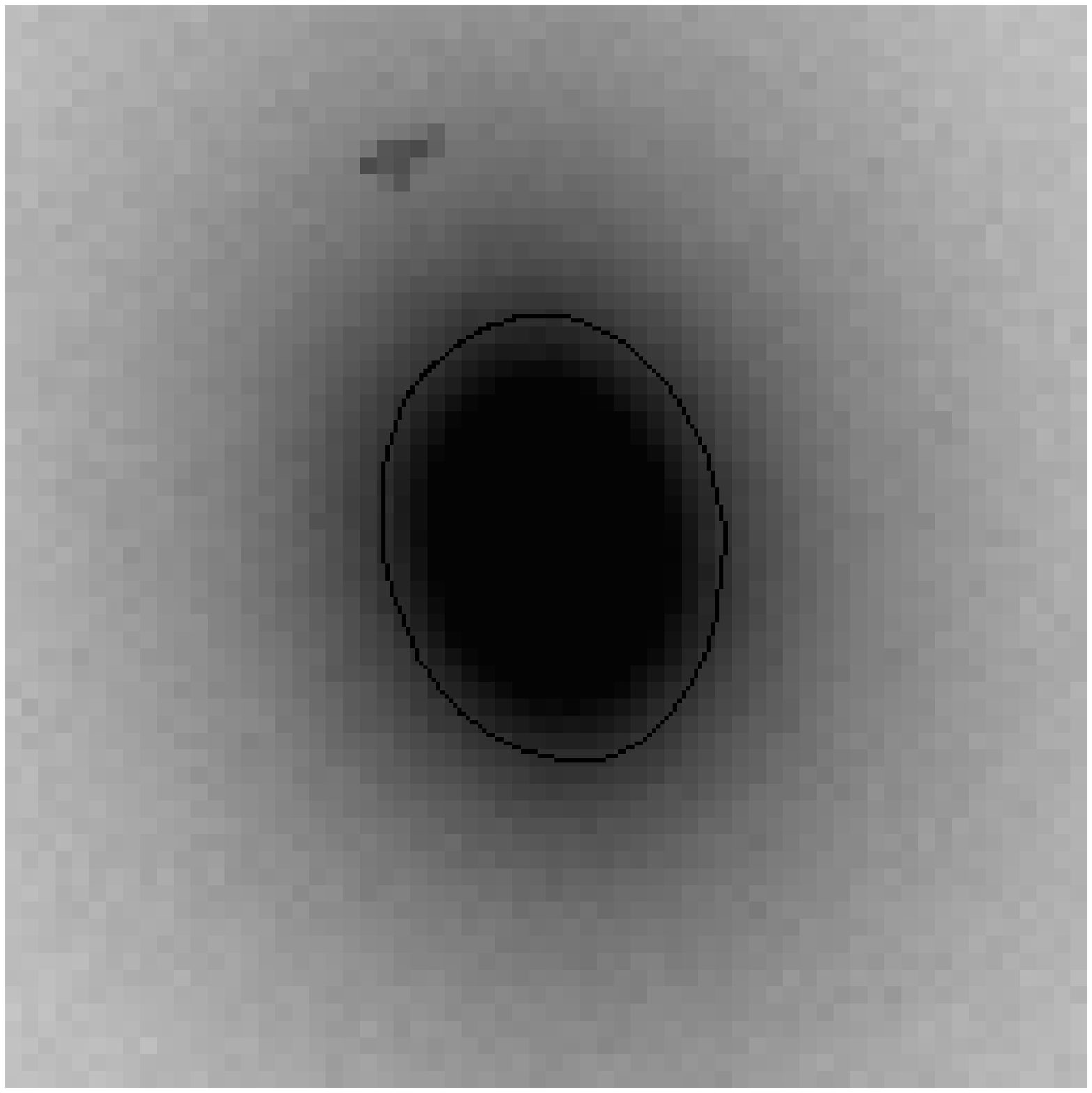,width=\dimen6,clip=}
\psfig{file=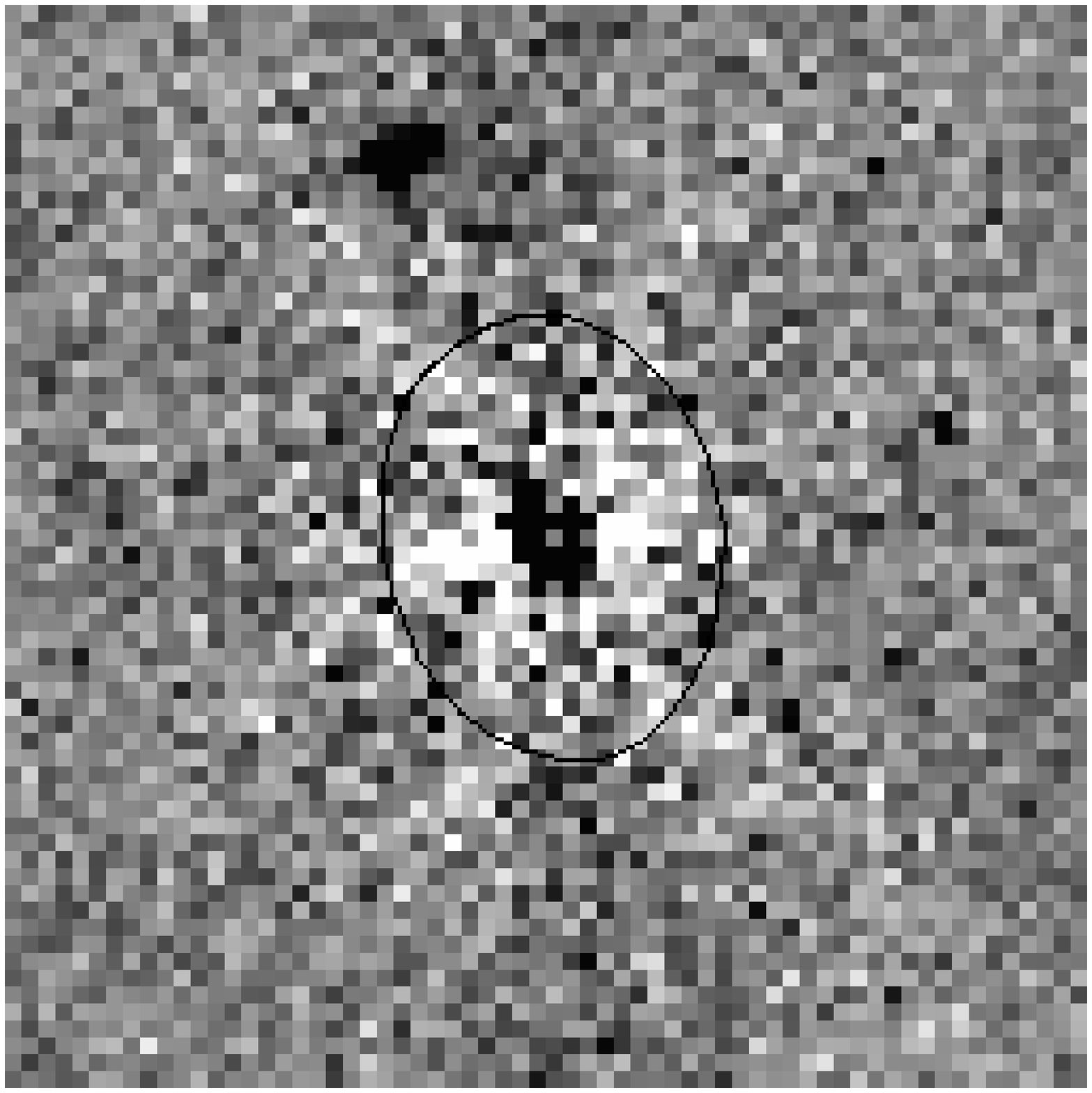,width=\dimen6,clip=}
}
\vskip4mm
\centerline{%
\psfig{file=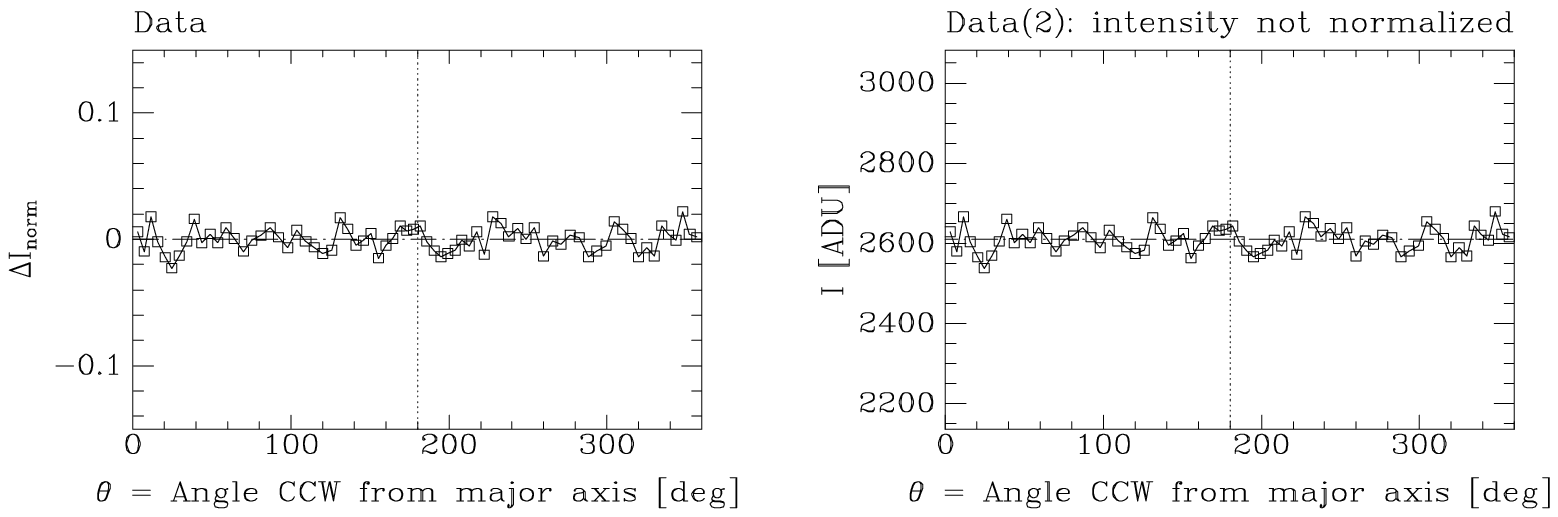,width=\dimen9,clip=}
}
\bcaption{5}
{Top: Original and residual images of the pure E galaxy, with a single ellipse
at $a \! = \! 2.6 \, {\rm kpc}$ ($r \! = \! 2.3 \, {\rm kpc}$) shown.
The images are 13 kpc on the side.
The `Data' plot shows $\dInorm$ versus $\theta$ (see text) along
the shown ellipse.
The `Data(2)' plot shows the same, except that the intensity $I$
is used rather than $\dInorm$.
Little structure is seen, consistent with the fact that all the
Fourier coefficients are close to zero, see Table~2.}
}
\endinsert

\vbox{%
\centerline{%
\psfig{file=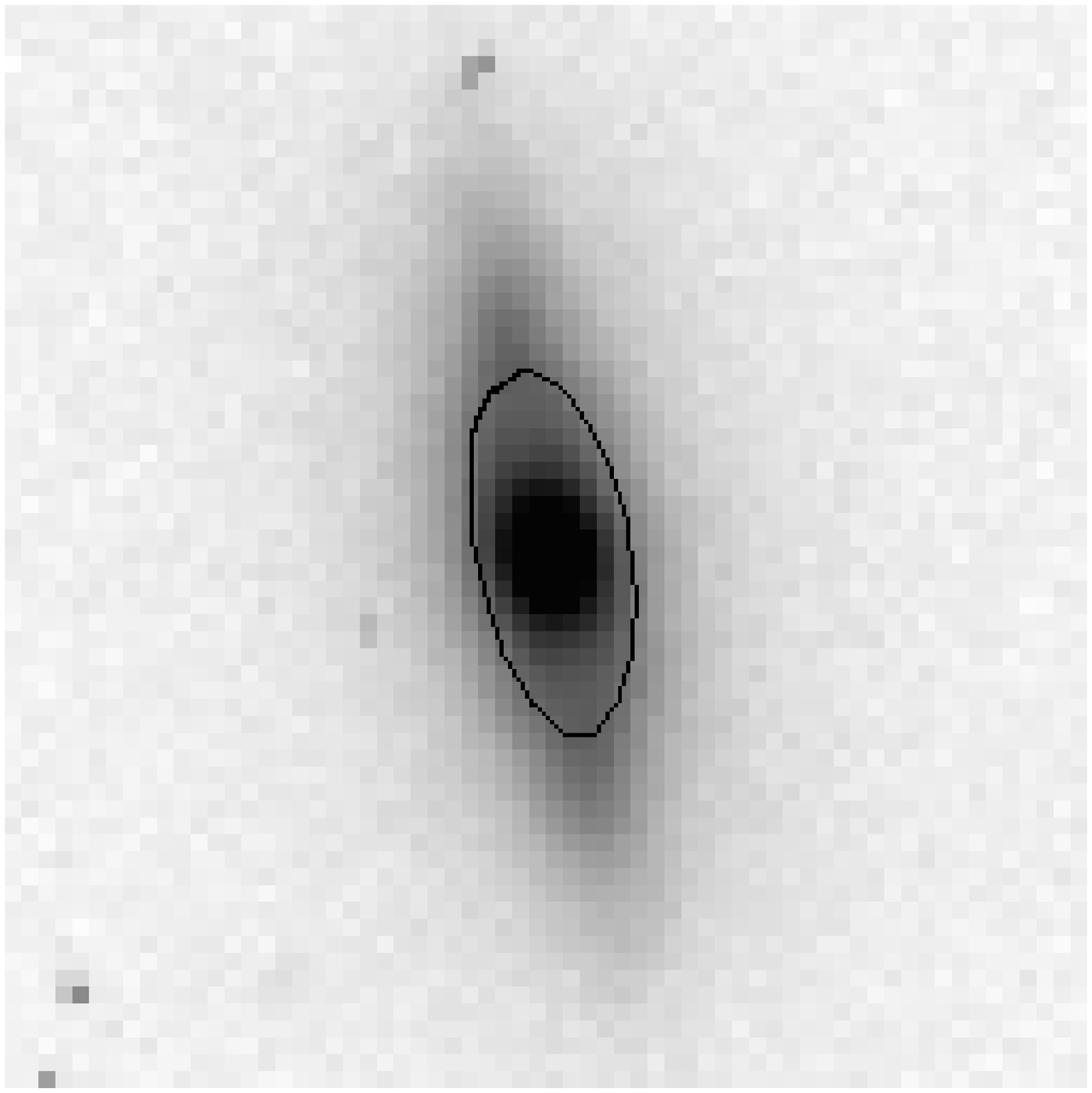,width=\dimen6,clip=}
\psfig{file=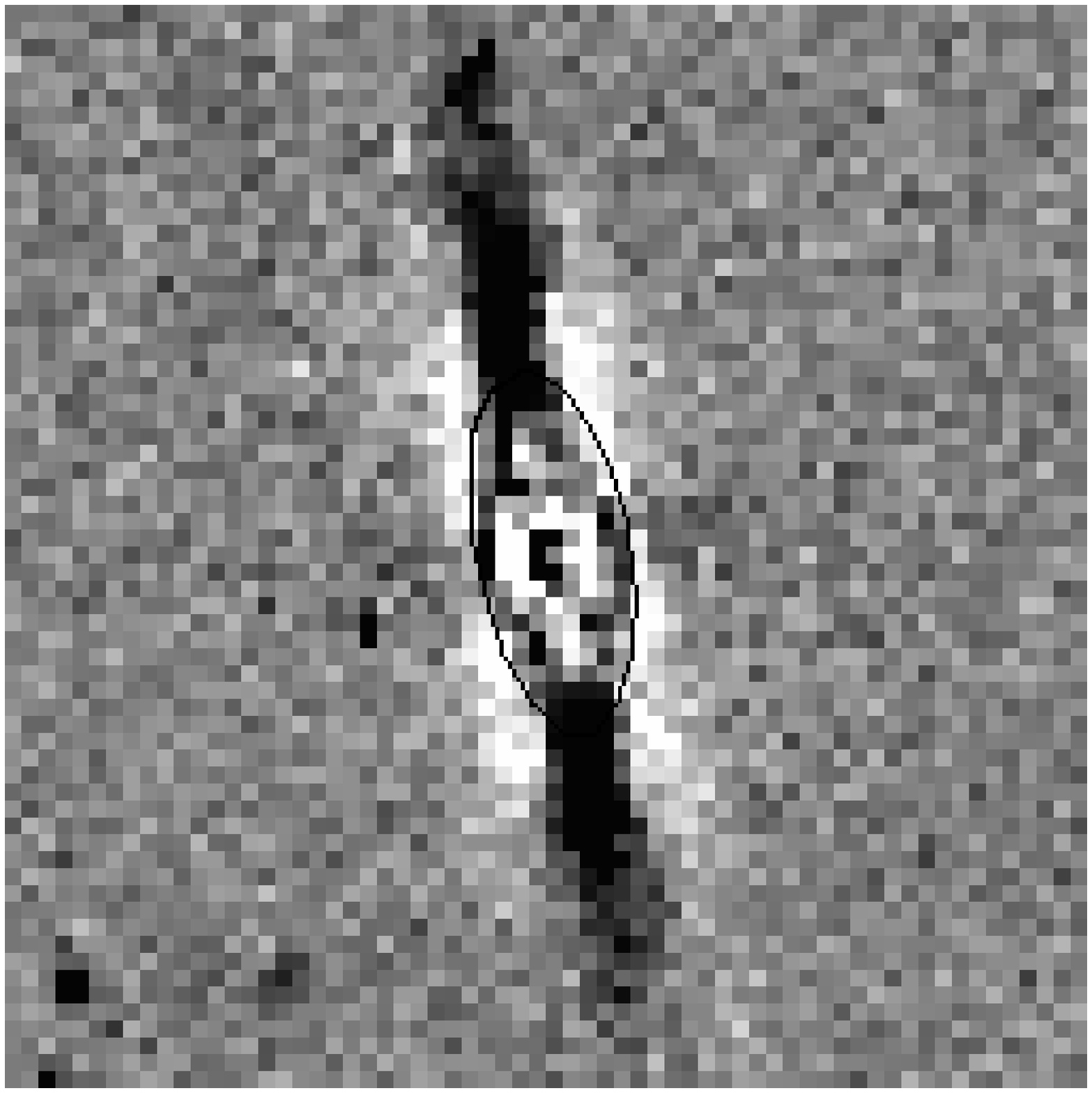,width=\dimen6,clip=}
}
\vskip4mm
\centerline{%
\psfig{file=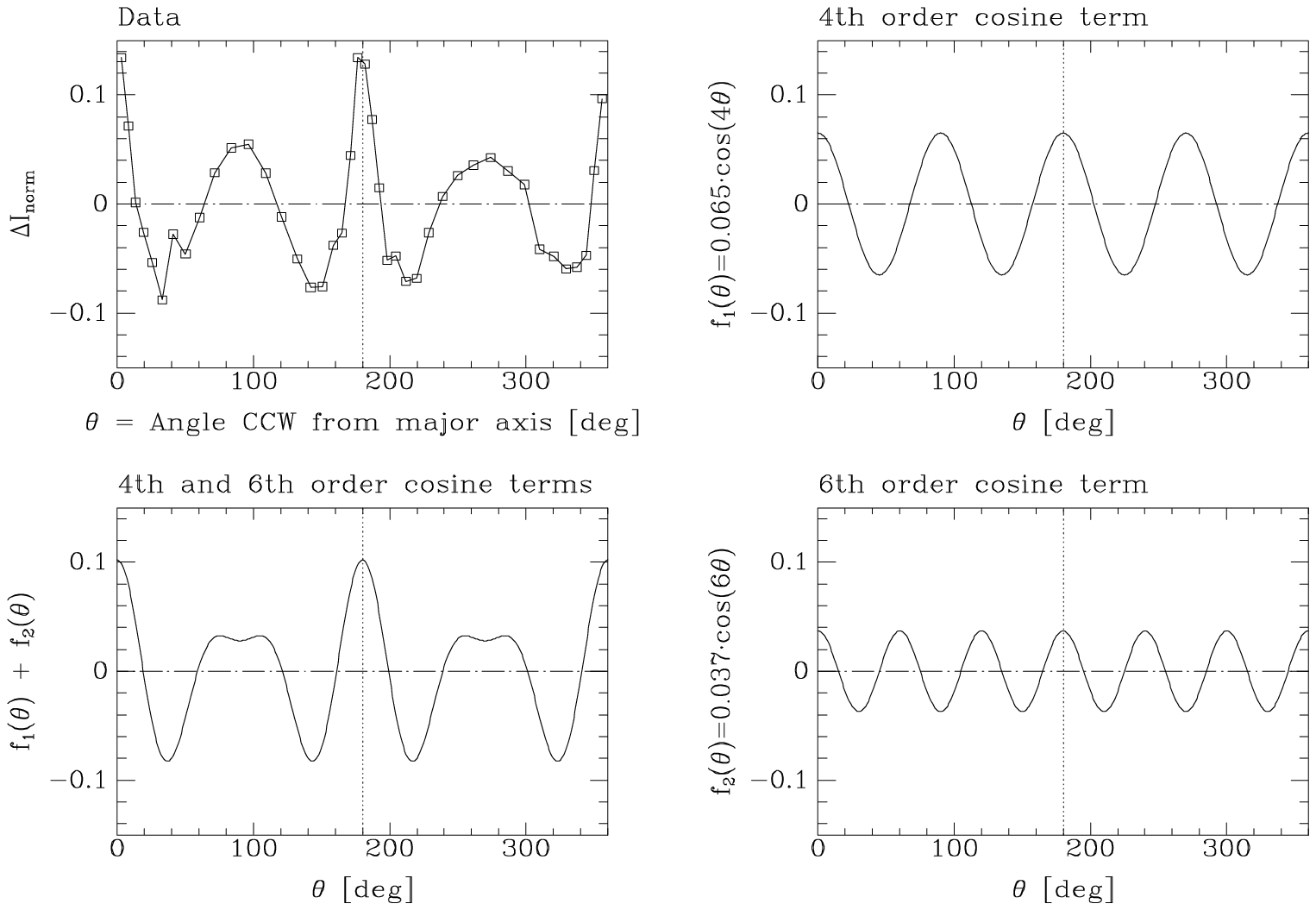,width=\dimen9,clip=}
}
\bcaption{6}
{Top: Original and residual images of the S0 galaxy, with a single ellipse
at $a \! = \! 2.2 \, {\rm kpc}$ ($r \! = \! 1.4 \, {\rm kpc}$) shown.
The images are 13 kpc on the side.
The `Data' plot shows $\dInorm$ versus $\theta$ along
the shown ellipse.
Substantial structure is seen.
The starting point of the curve ($\theta = 0^\circ$) corresponds to the
apogee of the ellipse that is closest to the top of the figure.
At that point the residual is positive (black).
The angle $\theta$ is measured counter-clockwise.
The two most dominant Fourier modes, the 4th and 6th order cosine terms,
cf.\ Table~2, are illustrated in the three other plots.
This example galaxy is a disky galaxy and as such has a positive $c_4$ 
(4th order cosine coefficient).}
}

\vbox{%
\centerline{%
\psfig{file=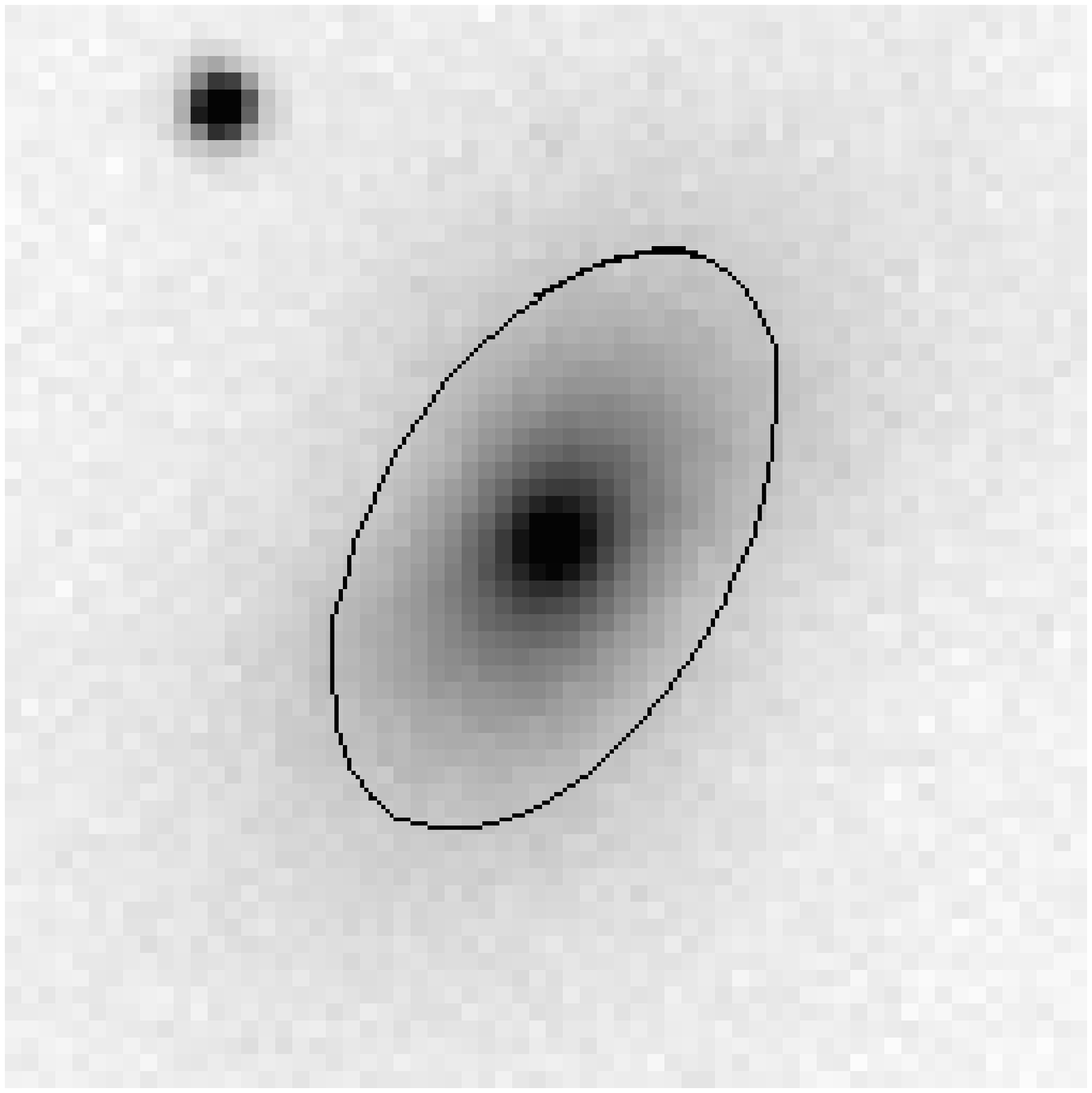,width=\dimen6,clip=}
\psfig{file=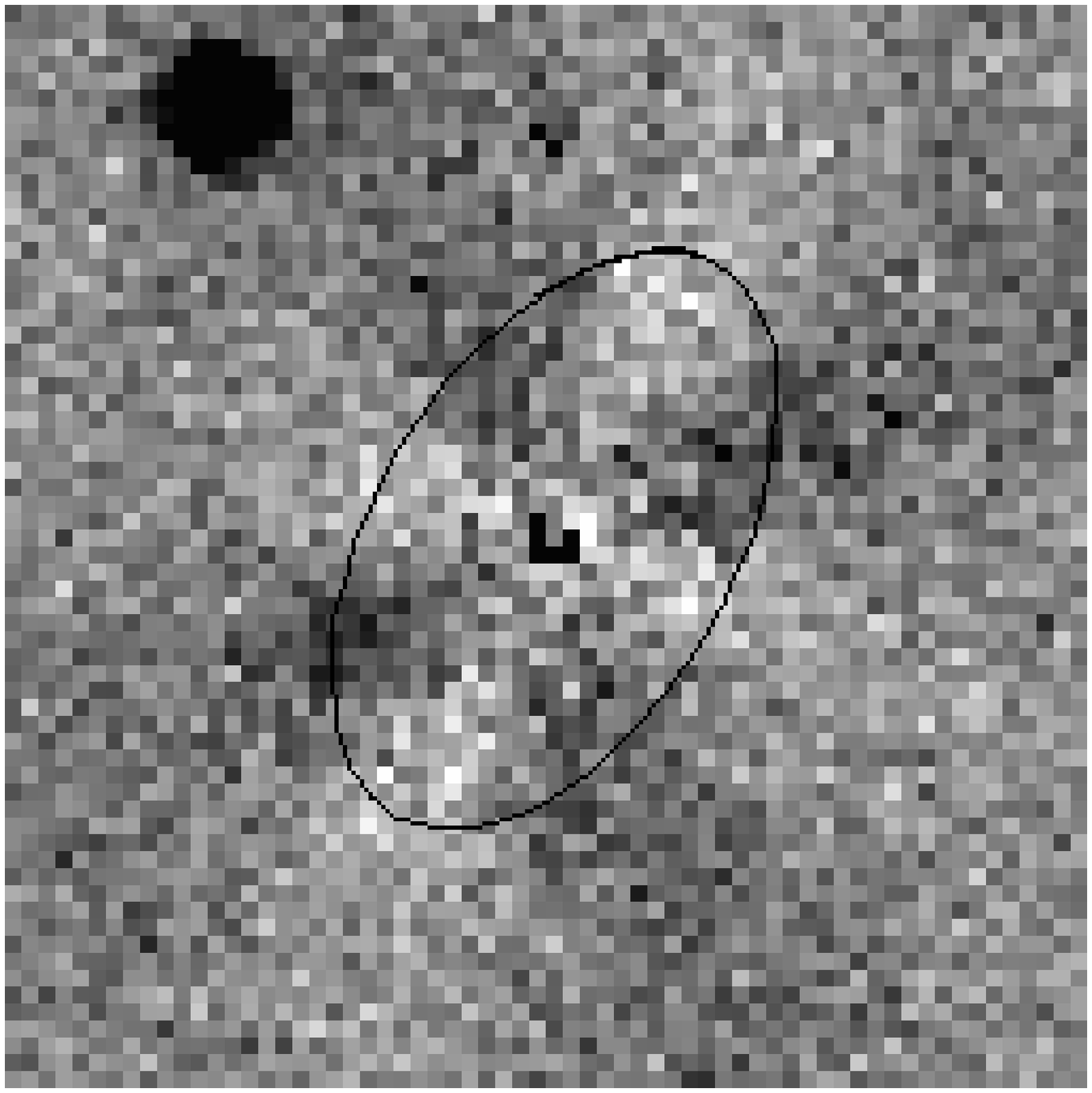,width=\dimen6,clip=}
}
\vskip4mm
\centerline{%
\psfig{file=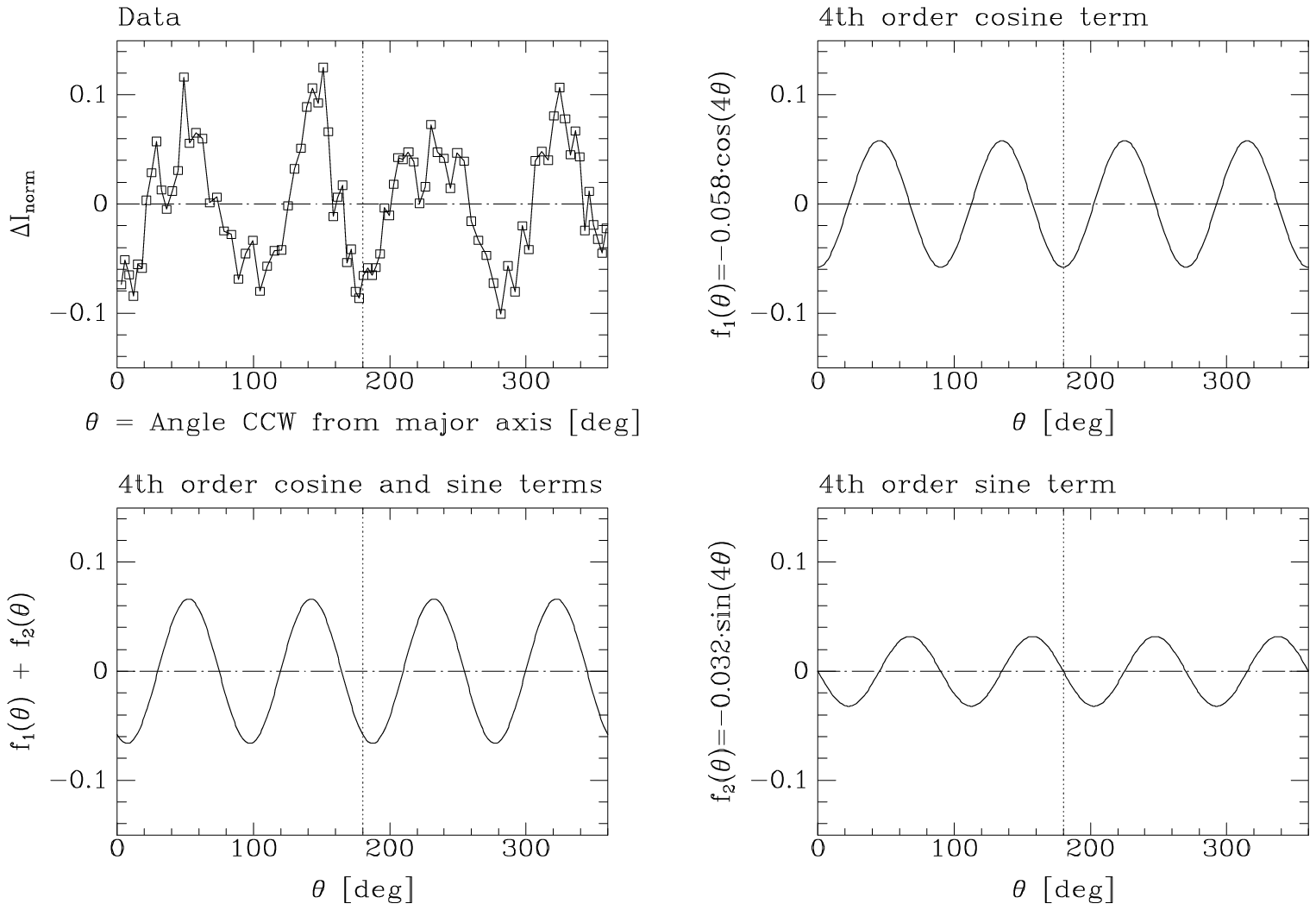,width=\dimen9,clip=}
}
\bcaption{7}
{Top: Original and residual images of the boxy galaxy, with a single ellipse
at $a \! = \! 3.8 \, {\rm kpc}$ ($r \! = \! 2.8 \, {\rm kpc}$) shown.
The images are 13 kpc on the side.
The `Data' plot shows $\dInorm$ versus $\theta$ along
the shown ellipse.
Substantial structure is seen.
The starting point of the curve ($\theta = 0^\circ$) corresponds to the
apogee of the ellipse that is closest to the top of the figure.
At that point the residual is negative (white).
The angle $\theta$ is measured counter-clockwise.
The two most dominant Fourier modes, the 4th order cosine and sine terms,
cf.\ Table~2, are illustrated in the three other plots.
Note how a boxy galaxy has a negative $c_4$ (4th order cosine coefficient).}
}

As can be seen, $\dInorm$ is the deviation in intensity $I$ from the mean
intensity at the given ellipse, $I_0$,
normalized by the equivalent radius $r$, and by the 
absolute value of the intensity gradient, $|{\rm d}I/{\rm d}r|$.
For comparison, we also show the intensity $I$ in Figure~5.
The reason for choosing this particular normalization will be 
described in the following.
One advantage of using the quantity $\dInorm$ in the plots
is that allows a direct comparison between the plots for
the three example galaxies.


To quantify how the {\it intensity\/} deviates from being constant
along the fitted ellipse,
the following Fourier series is fitted to the intensity $I(\theta)$
$$I(\theta) =
I_0 + \sum_{n=1}^{N} \bigl[A_n \sin(n\theta) + B_n \cos(n\theta)\bigr]
\enspace . \eqno(2)$$
$N$ is the highest order fitted.
$\theta$ is the angle measured counter-clockwise from the major 
axis of the ellipse.
The different Fourier modes, e.g.\ $\cos 4\theta$,
will be discussed below.

We are more interested in how the {\it isophote\/} deviates 
(in the radial direction) from the fitted ellipse.
Let $R_{\rm iso}(\theta)$ denote the distance from the center of the ellipse
to the isophote, and
let $R_{\rm ell}(\theta)$ denote the distance from the center of the ellipse
to the ellipse itself.
For a perfectly elliptical isophote, the difference between 
$R_{\rm iso}(\theta)$ and $R_{\rm ell}(\theta)$ would be zero
for all values of $\theta$.
We can Fourier expand the difference as
$$\Delta R(\theta) \equiv R_{\rm iso}(\theta) - R_{\rm ell}(\theta) =
\sum_{n=1}^{N} \bigl[A_n^\prime \sin(n\theta) + B_n^\prime \cos(n\theta)\bigr]
\enspace . \eqno(3)$$
$\Delta R(\theta)$ is the radial deviation of the isophote
from elliptical shape.
The {\it relative\/} deviation is more interesting,
so we take $\Delta R(\theta)$ relative to the size of the ellipse,
given by the equivalent radius $r$.
This relative radial deviation of the isophote from elliptical shape,
$\Delta R(\theta)/r$, is described by the
Fourier coefficients $A_n^\prime/r$ and $B_n^\prime/r$.
We will allocate new symbols for these quantities:
$s_n \equiv A_n^\prime/r$ and $c_n \equiv B_n^\prime/r$.

The ELLIPSE task calculates $A_n$ and $B_n$, but not
$A_n^\prime$ and $B_n^\prime$.
However, we are able to link the two sets of coefficients.
Consider a Taylor expansion to first order of $I(R)$ around $R_0$
$$I(R) = I(R_0) + {{\rm d}I \over {\rm d}R} (R - R_0) \enspace . \eqno(4)$$
Let $R$ be a point on the isophote and let $R_0$ be a point on the 
ellipse.
$I(R)$ is constant since $R$ is on the isophote.
The intensity on the ellipse $I(R_0)$ is not necessarily constant.
The intensity on the ellipse is also given by $I(\theta)$ (Equation 2).
We can identify the difference $(R - R_0)$ with $\Delta R(\theta)$
(Equation 3).
A suitable mean value of the gradient ${\rm d}I/{\rm d}R$
is ${\rm d}I/{\rm d}r$.
(The `effective intensity gradient' ${\rm d}I/{\rm d}r$ can be calculated as
the difference in intensity divided by the difference in equivalent radius
for two adjacent ellipses.)
By inserting Equation (2) and (3) in Equation (4) we find the following
relations hold for all $n$
$$A_n^\prime = {A_n \over |{{{\rm d}I \over {\rm d}r}}|} \enspace ,
\quad\quad\quad
  B_n^\prime = {B_n \over |{{{\rm d}I \over {\rm d}r}}|} \enspace ,
\eqno(5)$$
where we have used $-{\rm d}I/{\rm d}r = |{\rm d}I/{\rm d}r|$ since the
intensity gradient is negative.
 From the definitions of $s_n$ and $c_n$ given above,
we finally get
$$s_n = {A_n \over {r \cdot {\bigl|{{\rm d}I \over {\rm d}r}\bigr|}}}
\enspace ,
\quad\quad\quad
  c_n = {B_n \over {r \cdot {\bigl|{{\rm d}I \over {\rm d}r}\bigr|}}}
\enspace .
\eqno(6)$$
These definitions of the Fourier coefficients $s_n$ and $c_n$ are 
used in the lite\-rature by e.g.\
Franx, Illingworth \& Heckman (1989b);
J{\o}rgensen, Franx \& Kj{\ae}rgaard (1992);
J{\o}rgensen \& Franx (1994); and
J{\o}rgensen, Franx \& Kj{\ae}rgaard (1995).

Slightly different definitions are also in use.
Some authors, e.g.\
Bender \& M{\"o}llenhoff (1987);
Bender, D{\"o}bereiner \& M{\"o}llenhoff (1988);
Bender et al.\ (1989);
and
Nieto \& Bender (1989),
use $a_4/a$ for the 4th order cosine coefficient
(still for the radial deviation).
In our notation, this is equal to $B_4^\prime/a$.
The only difference between $c_4$ and $a_4/a$ is that
$c_4 \equiv B_4^\prime/r$ is taken relative to the equivalent radius $r$,
whereas
$a_4/a = B_4^\prime/a$ is taken relative to the semi-major axis $a$.
Thus, the two are related by
$$a_4/a = c_4 \, r/a = c_4 \sqrt{1-\epsilon} = c_4 \sqrt{b/a} \enspace ,
\eqno(7)$$
where we have used the known relations between $r$, $a$, $b$ and $\epsilon$.
For an apparently round galaxy (i.e.\ for $\epsilon = 0$),
$a_4/a$ is equal to $c_4$.

Yet another definition is used by e.g.\ Peletier et al.\ (1990).
These authors expand the intensity along the ellipse as
$$I(\theta) = I_0 \,
\Bigl(1 + \sum_{n=1}^{N} \bigl[S_n \sin(n\theta) + C_n \cos(n\theta)\bigr]\Bigr)
\eqno(8)$$
(note the upper case $C_n$).
When comparing with Equation (2) it is seen that
$I_0 S_n = A_n$ and $I_0 C_n = B_n$.
This means, for example, that $c_4$ and $C_4$ are related as
$$
C_4 = c_4 \, {r \over I_0} \cdot {\biggl|{{\rm d}I \over {\rm d}r}\biggr|}
    = c_4 \cdot {\biggl|{{\rm d}\log I \over {\rm d}\log r}\biggr|}
\enspace . \eqno(9)
$$
It also follows that $a_4/a$ is related to $C_4$ by
$$
a_4/a =
C_4 \cdot {\biggl|{{\rm d}\log I \over {\rm d}\log r}\biggr|}^{-1} \sqrt{b/a}
\enspace , \eqno(10)
$$
a relation used by Faber et al.\ (1997)
(but note the upper case $C_4$).

\midinsert
$$\vbox{\tabfont
\halign{
\huad \hfil#&           
\huad \hfil#&           
\huad \hfil#&           
\huad \hfil#\cr         
{\tbold Table 2.}
&\multispan{3}{\ Fourier coefficients for the three example galaxies at\hfil}\cr
&\multispan{3}{\ the particular ellipses shown in Figure~5, 6 and 7.\hfil}\cr
&\multispan{3}{\ The semi-major axis $a$ for these ellipses is given.\hfil}\cr
\tablerule
Coeff.\ & pure E          &                 S0 &             boxy E \cr
  & $a\!=\!2.6\,{\rm kpc}$ & $a\!=\!2.2\,{\rm kpc}$ & $a\!=\!3.8\,{\rm kpc}$ \cr
\tablerule
$c_3$ & $ 0.001 \pm 0.002$ & $-0.003 \pm 0.013$ & $ 0.002 \pm 0.008$ \cr
$s_3$ & $-0.001 \pm 0.002$ & $-0.000 \pm 0.014$ & $-0.002 \pm 0.008$ \cr
$c_4$ & $ 0.000 \pm 0.002$ & $ 0.065 \pm 0.013$ & $-0.058 \pm 0.006$ \cr
$s_4$ & $-0.004 \pm 0.002$ & $ 0.001 \pm 0.008$ & $-0.032 \pm 0.005$ \cr
$c_5$ & $ 0.000 \pm 0.002$ & $-0.003 \pm 0.008$ & $ 0.002 \pm 0.004$ \cr
$s_5$ & $ 0.001 \pm 0.002$ & $ 0.007 \pm 0.008$ & $-0.006 \pm 0.004$ \cr
$c_6$ & $ 0.003 \pm 0.001$ & $ 0.037 \pm 0.006$ & $-0.018 \pm 0.004$ \cr
$s_6$ & $-0.002 \pm 0.001$ & $-0.000 \pm 0.006$ & $ 0.012 \pm 0.004$ \cr
$c_7$ & $-0.000 \pm 0.001$ & $-0.001 \pm 0.006$ & $ 0.010 \pm 0.003$ \cr
$s_7$ & $-0.002 \pm 0.001$ & $-0.000 \pm 0.006$ & $-0.005 \pm 0.003$ \cr
$c_8$ & $ 0.004 \pm 0.001$ & $ 0.021 \pm 0.005$ & $-0.004 \pm 0.003$ \cr
$s_8$ & $-0.001 \pm 0.001$ & $ 0.003 \pm 0.005$ & $ 0.005 \pm 0.003$ \cr
\tablerule
}}
$$
\endinsert

The fitted values of the Fourier coefficients $c_n$ and $s_n$
for the example ellipses for the three example galaxies 
are listed in Table~2.
For the S0 galaxy, it is seen that the two numerically largest
coefficients are $c_4 = 0.065$ and $c_6 = 0.037$.
For the boxy galaxy the two numerically largest
coefficients are $c_4 = -0.058$ and $s_4 = -0.032$.

 From the definition of $\dInorm$ (Equation 1),
and from Equation (3) and (4) it is seen that $\dInorm = \Delta R(\theta)/r$,
i.e.\ $\dInorm$ measures the relative radial deviation 
of the isophote from elliptical shape.
This is also the quantity measured by $s_n$ and $c_n$ (indeed, they are the
Fourier coefficients of $\Delta R(\theta)/r$),
so plots of the Fourier modes $s_n \sin(n\theta)$ and $c_n \cos(n\theta)$
can be directly compared with the $\dInorm$ versus $\theta$ plot.
In Figure~6 and 7 we show the two most important Fourier modes 
and their sum for the S0 galaxy and the boxy E galaxy, respectively.
It is seen in both cases how the two most important modes
account for most of the structure.

\topinsert
\vbox{%
\centerline{%
\psfig{file=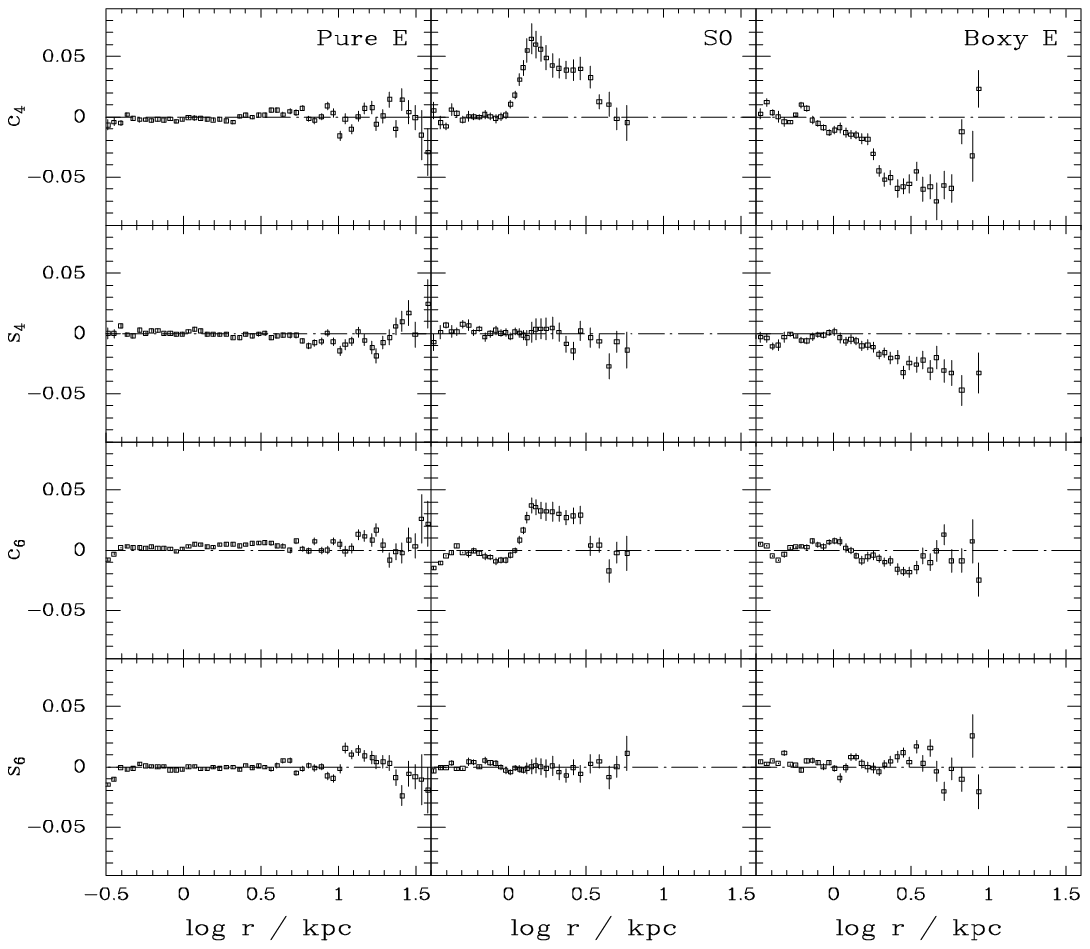,width=\dimen2,clip=}
}
\bcaption{8}
{Radial profiles of the Fourier coefficients $c_4$, $s_4$, $c_6$ and $s_6$
for the three example galaxies.
The plotted range in equivalent radii $r$ is 0.3--40 kpc (0.8$''$--102$''$).
Points with uncertainty larger than 0.025 are not plotted.}
}
\endinsert

An ellipse can be described by the first and second order Fourier coefficients.
Since the Fourier expansion is along the best-fitting ellipses,
the first and second order Fourier coefficients obtained there will be zero.

The cosine modes are symmetrical around the major axis ($\theta = 0^\circ$ or
$180^\circ$), whereas the sine modes are not, see Figure~7.
The odd cosine modes (e.g.\ 3rd order)
are not symmetrical around the {\it minor\/} axis
(see Figure 1a in Peletier et al.\ 1990).
The mode that is dominating for most E and S0 galaxies is the
4th order cosine.
Note from Figure~6 and 7 how
$c_4$ is an indicator of disky ($c_4 > 0$) or boxy ($c_4 < 0$)
isophotes aligned with the major axis
(e.g.\ Carter 1987; Bender et al.\ 1989; Peletier et al.\ 1990).

The radial profiles of the 4th and 6th order Fourier coefficients
are shown in Figure~8.

\subsubsection{2.2.1 The output from the ELLIPSE task}

The radial profiles determined by the ELLIPSE task are output to
an STSDAS table.
The most important columns are listed in Table~3,
along with their meaning in our notation.

The equivalent radius is not in the table, but it can be calculated as
{\tt SMA*sqrt(1-ELLIP)}.
The 5th to 8th order Fourier coefficients $s_n$ and $c_n$ can be
calculated as e.g.\ $c_6$ = {\tt BI6/(SMA*abs(GRAD))}.
This follows from Equation (6) since
$a \, {\rm d}I/{\rm d}a = r \, {\rm d}I/{\rm d}r$.
Note that the {\tt GRAD} column is ${\rm d}I/{\rm d}a$
(I. Busko 2000, private communication).

\midinsert
$$\vbox{\tabfont
\halign{
\huad #\hfil&           
\huad #\hfil\cr         
\multispan{2}{{\tbold Table 3.}~~Output columns from ELLIPSE\hfil}\cr
\tablerule
Column name    & Content in our notation \cr
\tablerule
{\tt SMA}      & $a$ \cr
{\tt INTENS}   & $I$ \cr
{\tt ELLIP}    & $\epsilon$ \cr
{\tt PA}       & PA (position angle) \cr
{\tt X0}       & $x$ (center of ellipse) \cr
{\tt Y0}       & $y$ (center of ellipse) \cr
{\tt GRAD}     & ${\rm d}I/{\rm d}a$ (not ${\rm d}I/{\rm d}r$) \cr
{\tt A}$n$     & $s_n$, $n$ = 3, 4 \cr
{\tt B}$n$     & $c_n$, $n$ = 3, 4 \cr
{\tt AI}$n$    & $A_n$, $n$ = 5, 6, 7, 8$^{\rm a}$ \cr
{\tt BI}$n$    & $B_n$, $n$ = 5, 6, 7, 8$^{\rm a}$ \cr
\tablerule
\multispan{2}{$^{\rm a}$~Only calculated with the following option set:\hfil}\cr
\multispan{2}{{\tt harmonics="5 6 7 8"}.\hfil}\cr
}}
$$
\endinsert

\subsection{2.3 Determination of magnitudes}

The intensity $I$ (in ADU) contains the signal from the galaxy
plus the sky background.
The sky background level can be determined in several ways.
One way is to identity {\it empty regions\/} in the image
and measure the level there. If the galaxy fills most of the image, this
can be difficult.
Another way is to fit a suitable analytical expression to outer part of
galaxy plus sky intensity profile obtained from the surface photometry.
This method has been used by e.g.\ J{\o}rgensen et al.\ (1992), 
fitting
$I_{\rm galaxy+sky}(r) = I_{\rm sky} + I_{\rm galaxy,0} \cdot r^{-\alpha}$,
with $\alpha$ = 2 or 3.

Magnitudes can be calculated from the sky subtracted intensity.
This can either be integrated magnitudes within a certain aperture
(elliptical or circular), or the surface brightness
at a given ellipse, $\mu(r)$.
By knowing the pixel scale of the CCD (in arcsec/pixel),
$\mu(r)$ can be expressed in units of mag/arcsec$^2$.
With the use of observed standard stars, the magnitudes and 
surface brightnesses can be transformed to a standard photometric system.

\subsection{2.4 Global parameters}

The surface photometry has produced radial profiles of a number of quantities.
It is desirable to condense these radial profiles to a few characteristic
numbers, the global parameters.

\subsubsection{2.4.1 Effective parameters}

Elliptical galaxies have surface brightness profiles that are well
approximated by the $r^{1/4}$ law (de Vaucouleurs 1948).
By fitting the aperture magnitudes to an $r^{1/4}$ growth curve,
the following two parameters can be derived:
\item {$\bullet$} $\re$: Effective radius, in arcsec
\item {$\bullet$} $\mue$: Mean surface brightness within $\re$,
                  in mag/arcsec$^2$
\vskip0.2mm
\noindent
The seeing needs to be taken into account (Saglia et al.\ 1993).

For galaxies with perfect $r^{1/4}$ profiles, the effective radius $\re$
is the {\it half-light\/} radius, i.e.\ the radius that encloses half of
the light from the galaxy.
Spiral galaxies
are better described by an exponential surface brightness profile 
than by an $r^{1/4}$ profile.

We can express the mean surface brightness in units of
$L_\odot/{\rm pc}^2$, where $L_\odot$ is the luminosity of the Sun in the
given passband (e.g.\ Gunn $r$).
We will call this quantity $\Ie$.
The relation is $\log \Ie = -0.4 (\mue - k)$,
where the constant $k$ is given by
$k = M_\odot + 5 \log (206265\,{\rm pc} / 10\,{\rm pc})$.
As is seen, the calculation does not involve the distance to the galaxy,
but only the absolute magnitude of the Sun in the given passband.
The calculation of $\re$ in kpc from $\re$ in arcsec, however,
does involve the distance to the galaxy.

With $\re$ as the half-light radius,
it follows that the total luminosity is given by
$L = 2 \pi \Ie^{} \re^2$.

\subsubsection{2.4.2 Global Fourier parameters}

As is seen from Figure~8, also the Fourier parameters vary with radius.
One way of getting a characteristic value of e.g.\ the $c_4(r)$ profile
is to take the extremum value. In case the profile does not have a clear
extremum, we can take the value at the effective radius.
We will use the symbol $c_4$ for this characteristic value of $c_4(r)$.

Another way to get a global Fourier coefficient
is to calculate an intensity weighted mean value as
$$
\barsn \equiv { \int_\rmin^\rmax I(r) \cdot s_n(r) \, {\rm d}r \over
                \int_\rmin^\rmax I(r) \, {\rm d}r } \enspace ,
\quad
\barcn \equiv { \int_\rmin^\rmax I(r) \cdot c_n(r) \, {\rm d}r \over
                \int_\rmin^\rmax I(r) \, {\rm d}r } \enspace ,
\eqno(11)
$$
where $\rmin$ is the radius where seeing effects are no longer important,
and $\rmax$ is the radius where the Fourier coefficients can no longer
reliably be determined (see J{\o}rgensen \& Franx 1994).

\subsubsection{2.4.3 Global ellipticities and colors}

As global ellipticity can be taken the extremum of the $\epsilon(r)$
profile, the value of $\epsilon(r)$ at the effective radius,
or the value of $\epsilon(r)$ at a certain isophote level,
e.g.\ $\mu = 21.85 \marc$ in Gunn $r$ as used by
J{\o}rgensen \& Franx (1994).


As global color, the color within the effective radius can be used.
By color is meant the difference between the magnitudes in two different
passbands, such as (B$-$V).
The color is always calculated as the magnitude in the passband with 
the shortest effective wavelength minus the magnitude in the passband
with the longest effective wavelength.
Thus, a large
value of the color means that the galaxy is red, and a small value means that
it is blue.
Another global parameter related to the color is the color gradient,
defined as $\Delta{\rm color} / \Delta\log r$, i.e.\ as the slope
of the color versus $\log r$ plot.

\vskip2truemm

\vfill\eject

\section{3. QUANTITATIVE MORPHOLOGY FOR E AND S0 GALAXIES}

CCD surface photometry as described in the previous section offers the
possibility of carrying out quantitative morphologic studies of E and 
S0 galaxies.
The traditional classification of these galaxies distinguishes between
E and S0 galaxies based on the presense of a disk (S0 galaxies) or no 
disk (E galaxies).
E and S0 galaxies are considered to belong to two separate classes of 
galaxies.

In the following we summarize the methods and results presented
by J\o rgensen \& Franx (1994, hereafter JF94). The results lead to the 
conclusion that E and S0 galaxies fainter than an absolute blue
magnitude of $\MBT =-22$ mag form one class of galaxies with a broad 
and continuous distribution of the relative luminosity of the disks.

\subsection{3.1. Sample properties and the data}

The sample used in the study by JF94 is a magnitude limited sample
of galaxies within the central square degree of the Coma cluster.
The sample is complete to an apparent magnitude in Gunn~$r$ of 
15.3 mag. 171 galaxies are included in the sample.
Because the sample is well-defined and complete it is possible based
on these data to draw conclusions regarding the E and S0 galaxies as 
a class.

CCD photometry was obtained of the full sample. Surface photometry 
for the galaxies was derived using GALPHOT
(J{\o}rgensen et al.\ 1992; 
Franx et al.\ 1989b). From the surface photometry, global parameters were
derived. The important parameters used in this study are summarized
in Table~4. $\barcfour$ and $\barcsix$ are intensity weighted mean
values of $\cfour$ and $\csix$, respectively (see JF94 and Section 2.4.2), while
$\cfour$ represents the extremum the $\cfour$ radial profile.
The intensity weighted parameters are less sensitive to small scale
features in the radial profiles and are therefore to be preferred
for studies of global properties.

\midinsert
$$\vbox{\tabfont
\halign{
\huad #\hfil&           
\huad #\hfil\cr         
\multispan{2}{{\tbold Table 4.}~~Surface photometry parameters\hfil}\cr
\tablerule
Parameter & Description  \cr
\tablerule
$\mT$        & Total magnitude in Gunn $r$      \cr
$\epsiso$    & Ellipticity at $\mu = 21.85 \marc$ in Gunn $r$ \cr
$\cfour$     & Extremum of the $\cfour(r)$ radial profile \cr
$\barcfour$  & Intensity weighted mean value of $\cfour(r)$ \cr
$\barcsix$   & Intensity weighted mean value of $\csix(r)$ \cr
\tablerule
}}
$$
\endinsert

\subsection{3.2. Morphology of the E and S0 galaxies}

Figure~9 shows the morphological parameters $\epsiso$, $\barcfour$
and $\barcsix$ as functions of the total magnitudes $\mT$ in Gunn $r$.
The spiral galaxies are also shown on these figures, but we will omit
them from the following discussion.
 From Figure~9 it is clear that the galaxies fainter than about
12.7 mag span the full range in morphological parameters. The E and
S0 galaxies cannot be separated into two classes of galaxies based
on one of these morphological parameters.
The difference in properties of E and S0 galaxies fainter and brighter
than $\mT = 12.7$ mag is striking.
This demarcation magnitude corresponds to a blue absolute magnitude
of $\MBT =-22$ mag.

In Figure~10 we show the distribution of the ellipticities, $\epsiso$,
of the galaxies. Figure~10b shows the cumulative frequency of
$\epsiso$ for the E galaxies and the S0 galaxies separately, as
well as the cumulative frequency of $\epsiso$ for the E and
S0 galaxies as one class.

\WFigure{9}{\psfig{figure=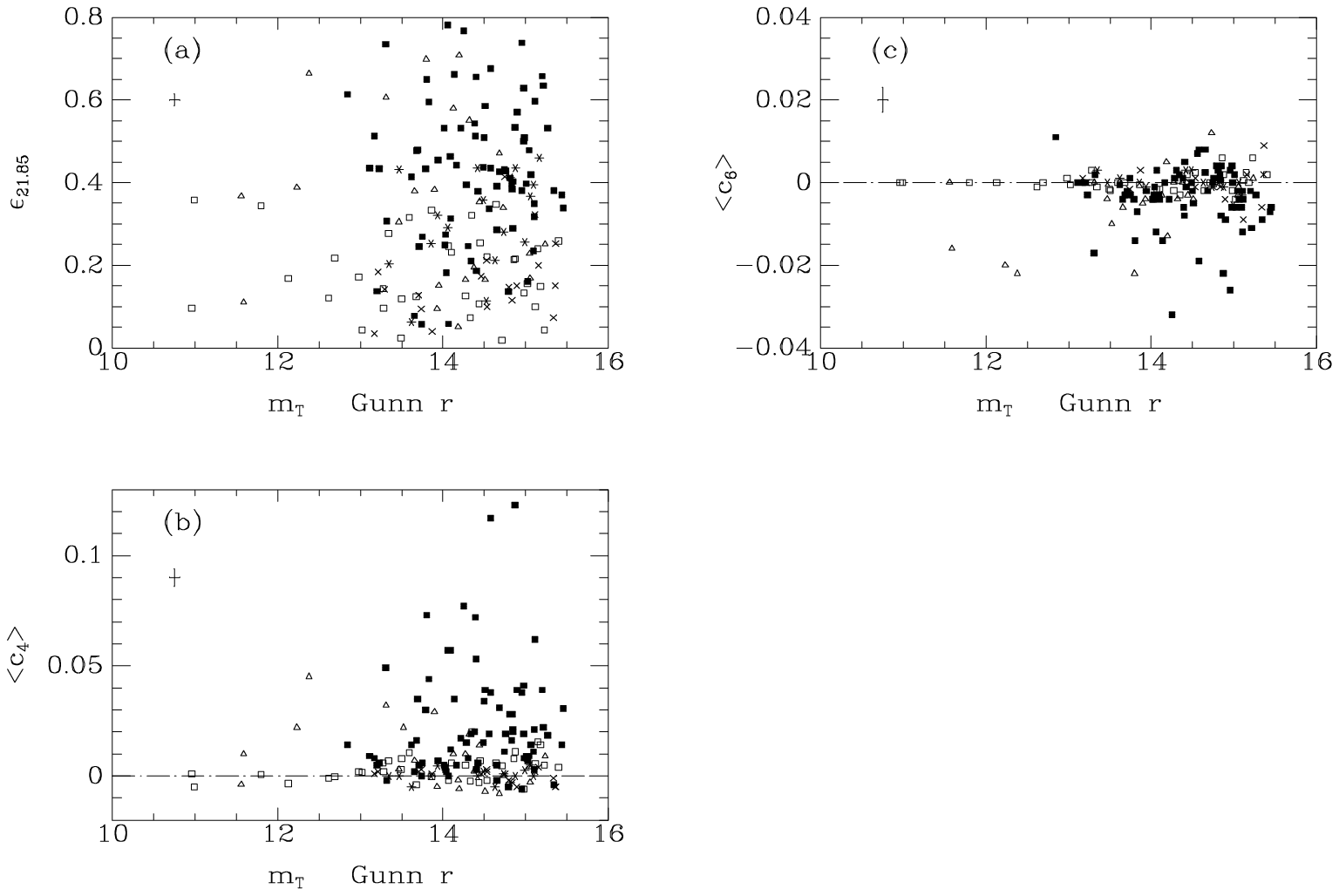,width=13truecm,angle=0,clip=}}
{Morphological parameters versus the total magnitudes.
Open boxes -- E galaxies with dominating regular $\cfour$-profiles; 
filled boxes -- S0  galaxies with dominating regular $\cfour$-profiles;
crosses -- E galaxies with irregular or non-dominating
$\cfour$-profiles; 
stars -- S0 galaxies with irregular or non-dominating 
$\cfour$-profiles;
triangles -- spirals.
The $\cfour$-profiles are considered non-dominating if $|\barcfour |$
is more than one sigma smaller than the absolute value of one
of the other intensity weighted mean coefficients.
Typical measurement errors are given on the panels.
Galaxies with uncertainty on $\cfour$ respectively $\barcfour$ larger
than 0.02 are not plotted.
The six brightest galaxies have small Fourier coefficients and 
$\epsiso < 0.4$. Other dependence on $\mT$ is not seen.
(From JF94.)
}

If the E galaxies and the S0 galaxies form two separate classes, then
we expect that their $\epsiso$ distributions can be modeled 
independently. The simplest assumption is that the galaxies are 
randomly oriented in space and have some simple distribution
of the {\it intrinsic\/} ellipticities.
JF94 attempted to fit the $\epsiso$ distributions with intrinsically
uniform ellipticity distributions and with intrinsic ellipticity
distribution that were Gaussians. The fit was done by maximizing the
probability that the data was drawn from the model as reflected by
a Kolmogorov--Smirnov (K--S) test (e.g.\ Press et al.\ 1992). 
The K--S test gives the probability
that a data distribution is drawn from a model distribution (or
another data distribution) based on the maximum difference between
the cumulative frequencies of the two distributions, as it is illustrated
in Figure~10b.
JF94 found that the $\epsiso$ distribution of the E galaxies could
be fitted satisfactory with either a uniform or a Gaussian intrinsic
distribution of the ellipticities. 
Both of these models resulted in probabilities of 84 per cent or larger.
However, the $\epsiso$ 
distribution of the S0 galaxies could not be fitted satisfactory.
Both intrinsic distributions have probabilities of only 11 per cent.
The $\epsiso$ distribution of the S0 galaxies lack apparently
round galaxies, see Figure~10b.
JF94 also find that the $\epsiso$ distribution of the E and S0 galaxies
treated as one class can be fitted satisfactory by either a uniform
or a Gaussian intrinsic distribution of the ellipticities.

\midinsert
\vbox{%
\centerline{\psfig{figure=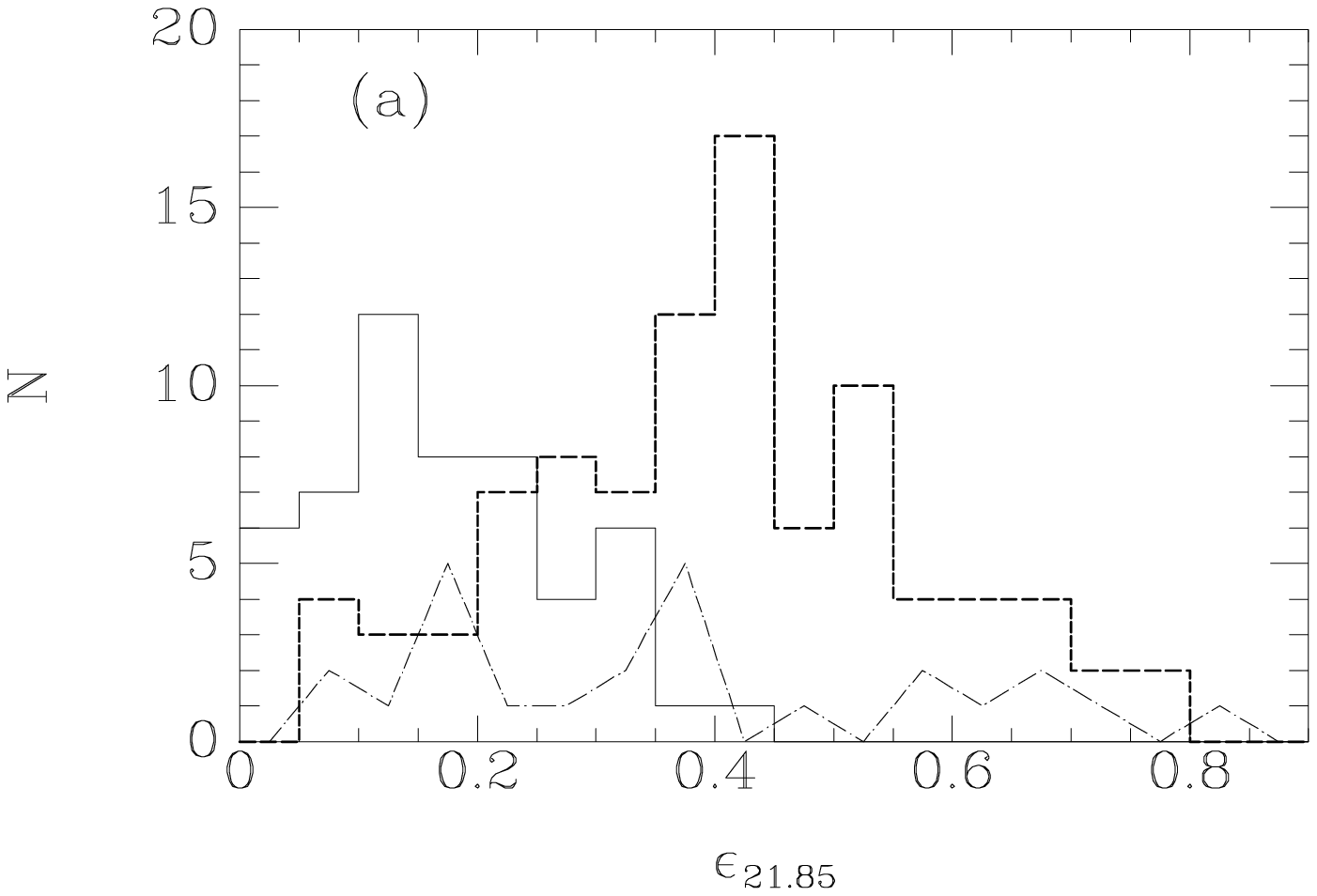,width=6.5truecm,angle=0,clip=}
\psfig{figure=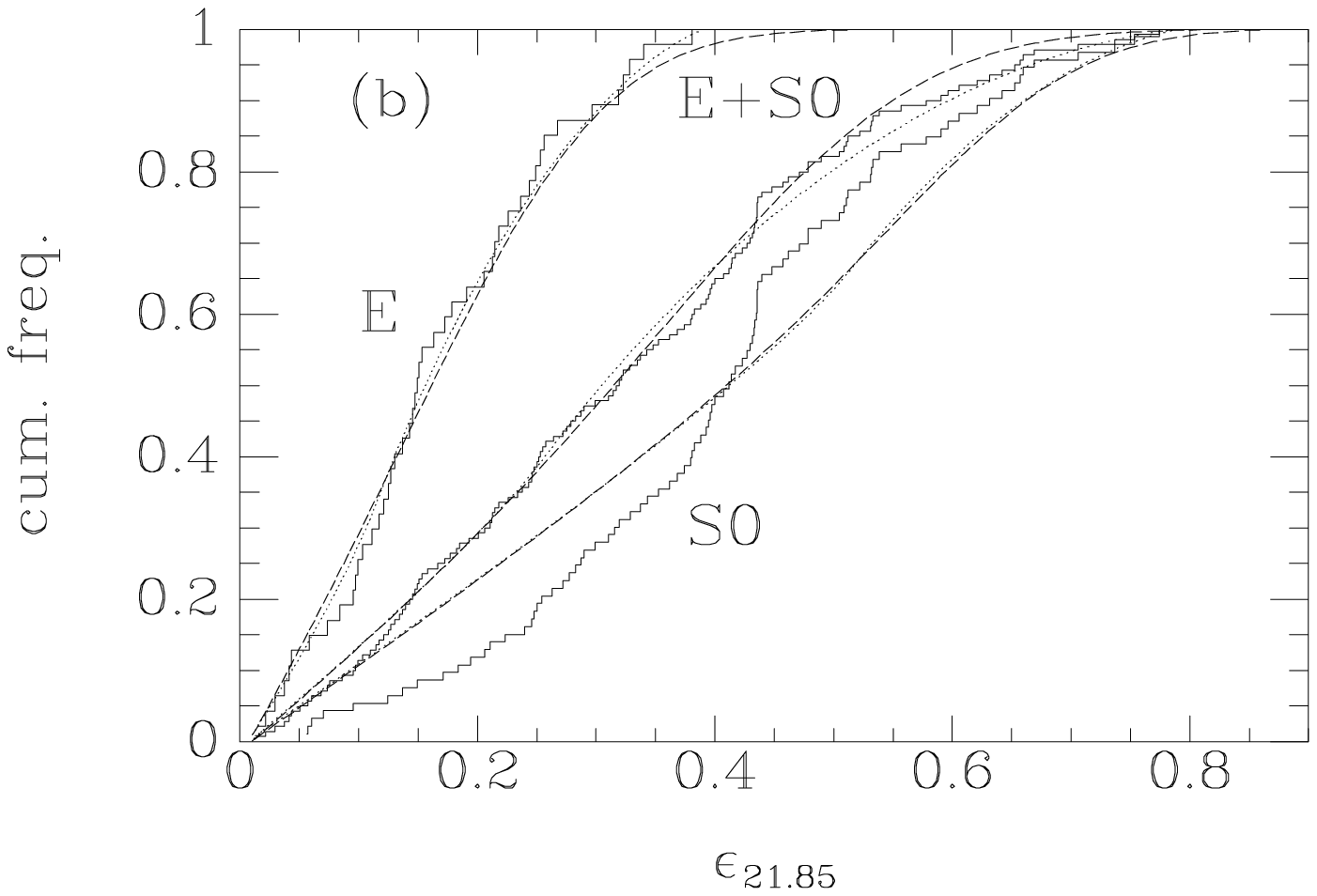,width=6.5truecm,angle=0,clip=}}
\vskip2truemm
\bcaption{10}
{Distributions of the apparent ellipticities.
(a) Solid line -- E galaxies, dashed line -- S0 galaxies,
Dashed-dotted line -- spirals.
(b) Solid lines -- E, S0 galaxies, and E and S0 galaxies together.
The six brightest E galaxies are excluded.
Dotted lines -- best fitting uniform distributions.
Dashed lines -- best fitting Gaussian distributions.
(From JF94.)
}}
\endinsert

These results show that some (maybe all) of the E galaxies
fainter than $\mT = 12.7$ mag must be S0 galaxies seen face-on. 
When the galaxies are seen face-on a disk is more difficult to 
detect (as also noted by Rix \& White 1990), and the galaxies are
mostly classified as E galaxies even when a disk is present.

\vskip1truecm

\subsection{3.3 The relative disk luminosities}

The next task is to determine the relative disk luminosities, 
$\LD /\Ltot$, of the galaxies. $\LD /\Ltot$ is the luminosity of the
disk relative to the total luminosity of the galaxy.
JF94 showed that two of the morphological parameters can be used
together to derive $\LD /\Ltot$ if a simple model for the disk and
the bulge is assumed.

\WFigure{11}{\psfig{figure=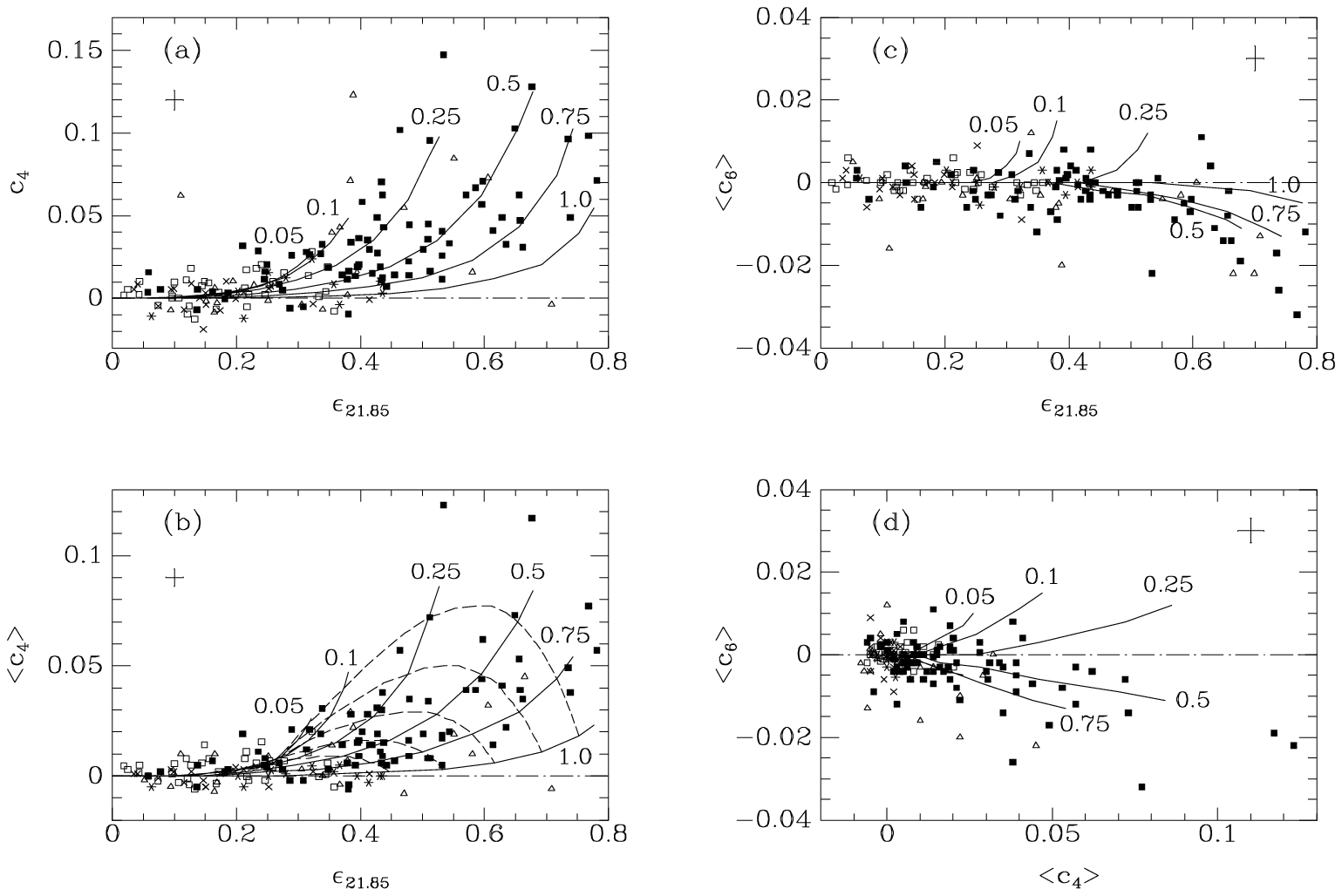,width=13truecm,angle=0,clip=}}
{ Fourth and 6th order Fourier coefficients plotted against
ellipticity and versus each other.
Data symbols as in Figure~9.
Galaxies with uncertainty on $\cfour$ respectively $\barcfour$ larger
than 0.02 are not plotted.
Typical measurement errors are given on the panels.
The models with $\aeB = \aeD$ are overplotted.
The curves are labeled with $\LD /\Ltot$.
The dashed lines on (b) mark the inclinations where $\cos i = 0.1$,
0.2, 0.3, 0.4, and 0.5
(i.e.\ $i = 84^\circ$, $78^\circ$, $73^\circ$, $66^\circ$, and $60^\circ$),
from top to bottom.
No dashed line is shown for the $i = 90^\circ$ (i.e.\ edge-on) models,
but this line would be above the $i = 84^\circ$ dashed line,
connecting the end points of the solid lines.
It is seen that for a given $\LD /\Ltot$ the highest inclination gives
the largest $\barcfour$.
The non-zero coefficients for the pure disk-model are due to the
inclusion of seeing effects in the models.
(From JF94.)
}

Figure~11 illustrates this technique for the Coma cluster sample.
JF94 constructed models consisting of a bulge with an $r^{1/4}$
profile and a disk with an exponential profile. The bulge is assumed
to have an intrinsic ellipticity of 0.3, while the intrinsic 
ellipticity of the disk is assumed to be 0.85.
JF94 tested models for both equal major axis of the two 
components, $\aeB = \aeD$, and for $\aeB = 0.5\aeD$. The results are 
not significantly different, so here we will concentrate on 
the $\aeB = \aeD$ models.
JF94 convolved the model images of bulge plus disk with a representative
seeing and then analyzed the model images in the same way as done for
the data. 
Models with relative disk luminosities $\LD /\Ltot$ between zero 
(no disk) and one (all disk) were constructed. Further, the inclination
was varied between face-on (small inclination, cos $i$ = 1) and
edge-on (large inclination, cos~$i$ = 0) in steps of 0.1 in cos $i$.

The results in terms of morphological parameters are shown
on Figure~11. The models reproduce the general variation of $\cfour$,
$\barcfour$ and $\barcsix$ with ellipticity. They also span
reasonably well the section of the $\barcfour$--$\barcsix$ diagram
covered by the data.
Models of the kind used by JF94 cannot reproduce boxy isophotes of
the galaxies. However, the Coma cluster sample contains only two
galaxies fainter than $\mT = 12\fm 7$ that have $\barcfour$ 
significantly smaller than zero.
For galaxies with ellipticities larger than 0.3 and $\barcfour$ larger 
than 0.007, the models are well separated in $\barcfour$ 
versus $\epsiso$, see Figure~11b.
Thus, these two parameters can be used to derive the relative disk 
luminosities, $\LD /\Ltot$, and the inclinations, $i$, of the galaxies.

\midinsert
\vbox{%
\centerline{\psfig{figure=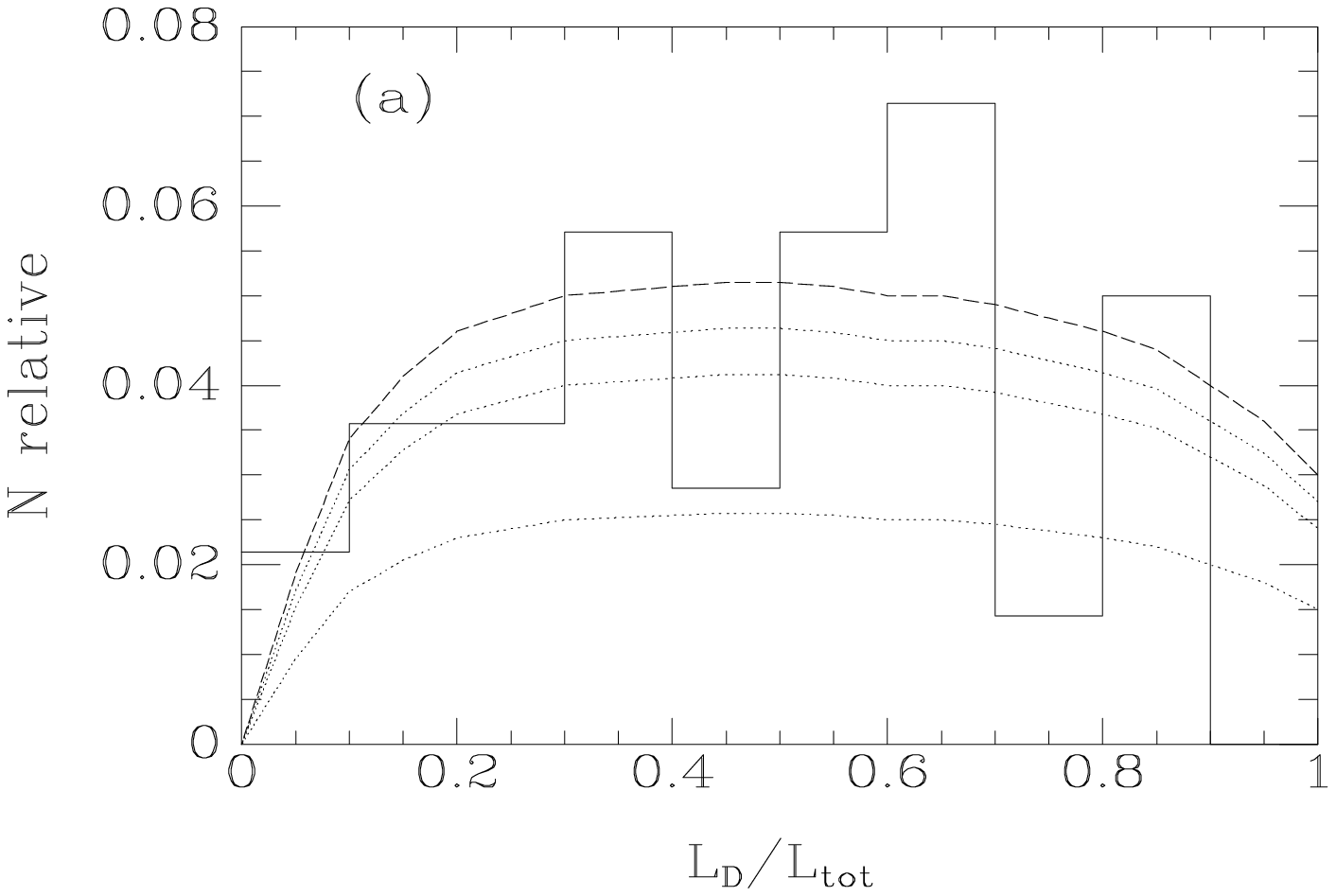,width=6.5truecm,angle=0,clip=}
\psfig{figure=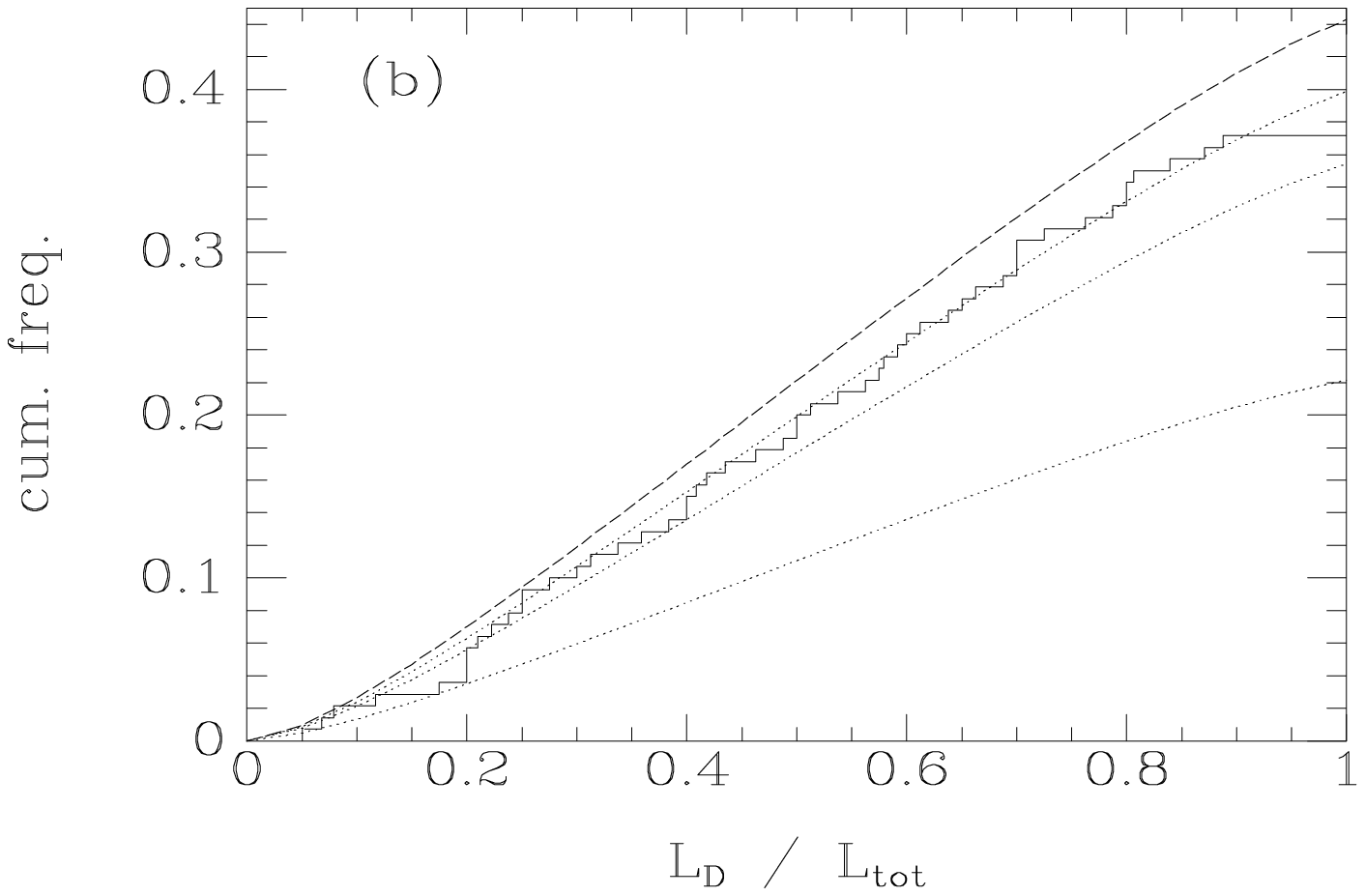,width=6.5truecm,angle=0,clip=}}
\vskip2truemm
\bcaption{12}
{Relative (a) and cumulative (b) frequency for
the relative disk luminosities.
Solid line -- determinations from the $\epsiso$--$\barcfour$ diagram.
The distribution has been normalized with the total number of
galaxies fainter than $\mT = 12\fm 7$.
Dashed line -- model prediction
for a uniform intrinsic distribution, corrected for
the incompleteness due to the limits enforced on $\epsiso$
and $\barcfour$.
Dotted lines -- model predictions for a uniform intrinsic distribution
plus a fraction of diskless galaxies.
Models for fractions of 0.1, 0.2, and 0.5 are shown.
A higher fraction of diskless galaxies moves the curve for the
normalized distribution downwards in both (a) and (b).
(From JF94.)
}}
\endinsert

Figure~12 shows the resulting distribution of $\LD /\Ltot$. It was 
possible to derive $\LD /\Ltot$ for 52 of the E and S0 galaxies in the
sample. The cumulative frequency shown on Figure~12b is normalized
to the 140 E and S0 galaxies fainter than $\mT =12.7$ mag.
Overplotted on Figure~12 are models of the distribution of $\LD /\Ltot$.
These models are uniform distributions with some fraction of diskless
galaxies added. The best fitting model is a uniform distribution of
$\LD /\Ltot$ between zero and one with an additional 10 per cent 
diskless galaxies.
JF94 find that the resulting distributions of
inclinations, $\epsiso$ and $\barcfour$
for this model also fit the observed distributions of these parameters.

\subsection{3.4 Conclusions from JF94}

The results presented by JF94 illustrate the strengths of studies
of quantitative morphology based on surface photometry for statistically
well-defined and complete samples.
JF94 were able to show that the E and S0 galaxies in the Coma cluster
(fainter than $\mT = 12.7$ mag in Gunn $r$) form one class of galaxies
with a broad (most likely uniform) distribution of $\LD /\Ltot$.
This result contradicts the traditional classification of these
galaxies into two separate classes. It also provides constraints
for models for morphological evolution of galaxies, since the
end-result for the Coma cluster represents one scenario that the
models need to reproduce.

\vskip2truemm

\section{4. STELLAR POPULATION MODELS}

Stellar population models are tools for interpreting the integrated 
light, such as the colors, observed from galaxies.
Ideally, we want to determine the mix of stars that give rise to
the observations.
This problem is usually underconstrained, so it is necessary 
to make some assumptions regarding how the numbers of different 
types of stars are related.
Here we will consider so-called single-age single-metallicity models,
also known as single stellar population (SSP) models.
In these models, all the stars are formed at the same time,
with distribution in mass given by the chosen
initial mass function (IMF), and with identical chemical composition.

SSP models are based on the following ingredients.
First, theoretical stellar isochrones are needed.
Isochrones are loci in the theoretical HR-diagram $(\log T_{\rm eff},\log L)$
for a stellar population of a given age and chemical composition.
Second, a conversion is needed between the theoretical parameters 
of $T_{\rm eff}$, $L$,
$\log g$ (surface gravity) and the metallicity [M/H] to the 
observable parameters such as colors.
This conversion can be either empirical or theoretical.
The empirical conversion is based on observations of individual stars
in our Galaxy
for which the `theoretical parameters' can be inferred, and the
observable parameters measured.
The theoretical conversion is based on model atmospheres and synthetic spectra.
Third, the IMF has to be specified.

An example of SSP models are those presented by Vazdekis et al.\ (1996).
These models use the isochrones from the Padova group (Bertelli et al.\ 1994).
The conversion from the theoretical to the observable parameters is empirical.
Models are presented for several different IMFs.
One IMF is a constant below 0.2 $M_\odot$,
a Salpeter (1955) IMF above 0.6 $M_\odot$,
and a spline in the interval 0.2--0.6 $M_\odot$.
The models that we use in Section 5 are based on this IMF.

When the IMF has been specified, the models have only
two parameters: age and metallicity.
The metallicity can be expressed either as the mass fraction in heavier
elements, $Z$, or as the metal abundance
[M/H] $\equiv \log(Z/Z_\odot)$ (with $Z_\odot = 0.02$).
The models have solar abundance ratios, while
this may not be the case for all galaxies.

\WFigure{13}{%
\psfig{file=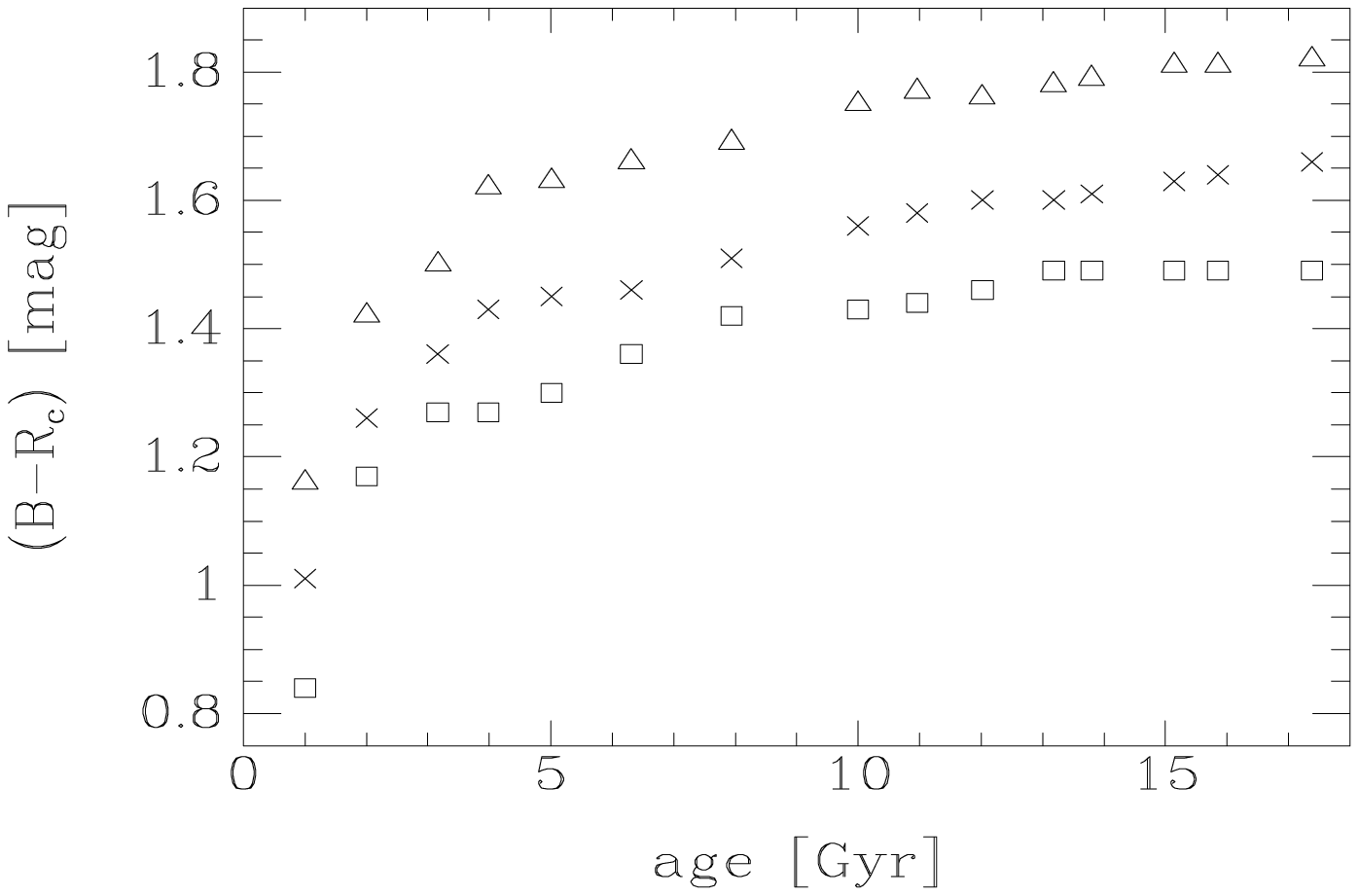,width=\dimen3,clip=}
}
{Example of the predictions from the Vazdekis et al.\ (1996) models:
(B$-$R$_{\rm c}$) color as function of age (the $x$-axis) and metallicity
(the different symbols).
Triangles -- ${\rm [M/H]} \! = \!  0.4$;
crosses   -- ${\rm [M/H]} \! = \!  0.0$;
boxes     -- ${\rm [M/H]} \! = \! -0.4$.
It is seen that the stellar populations get redder with higher age and
higher metallicity.}

As an example, 
the model predictions for the (B$-$R$_{\rm c}$) color is given in
Figure~13. It is seen that the color depends both on age and metallicity.
As will be illustrated in the next section,
optical colors depend roughly in the same way on age and metallicity.
Therefore, in a plot of one optical color versus another optical color
the model lines of constant age will be almost on top of the model lines
of constant metallicity.
The effects of age and metallicity cannot be disentangled in
such a diagram.
This is known as the {\it age--metallicity degeneracy\/} (e.g.\ Worthey 1994).

It should be noted that real galaxies are not necessarily
single stellar populations. For example, a galaxy could have experienced
a second star formation event.
Therefore, when SSP model predictions are compared with data for real galaxies
to determine the age and the metallicity, the resulting
ages and metallicities are luminosity weighted mean values.
Further, dust can also cause a stellar population to appear red.
This can be an additional complication in determining ages and metallicities.

\vskip2truemm

\vfill\eject

\section{5. COLOR RELATIONS}

The E and S0 galaxies follow very well-defined relations between
the optical colors and the total magnitudes. Figure~14 shows the
optical color-magnitude relation for the Coma cluster.
The figure includes all objects detected in a field covering the
central 75 arcmin $\times$ 80 arcmin of the cluster.
The data were obtained with the McDonald Observatory 0.8-meter
Telescope equipped with the Prime Focus Camera.

The color-magnitude relation is well-defined and has a very
low scatter for (E and S0) galaxies brighter than about 
$R_{\rm c} = 17$ mag, see Figure~14.
For galaxies fainter than $R_{\rm c} = 17.5$ mag only sparse
redshift information is available, and many of these faint
galaxies may be background galaxies.

The color-magnitude relation is thought to be primarily a result
of differences in metallicity as a function of luminosity.
However, recent results based on spectroscopy show that both the 
mean ages and the mean metallicities varies for E and S0 galaxies 
at low redshifts (e.g.\ Worthey, Trager \& Faber 1995; J\o rgensen 1999).
Thus, there is a need for interpreting the color-magnitude relation
within these recent results in order to achieve an self-consistent 
interpretation of the spectroscopic and the photometric results.
The work by Kauffmann \& Charlot (1998) represents one of
the only attempts to model the color-magnitude relation and
relations involving spectroscopic information in a consistent 
manner. In the models by Kauffmann \& Charlot both age and metallicity
varies with the luminosity.

\WFigure{14}{\psfig{figure=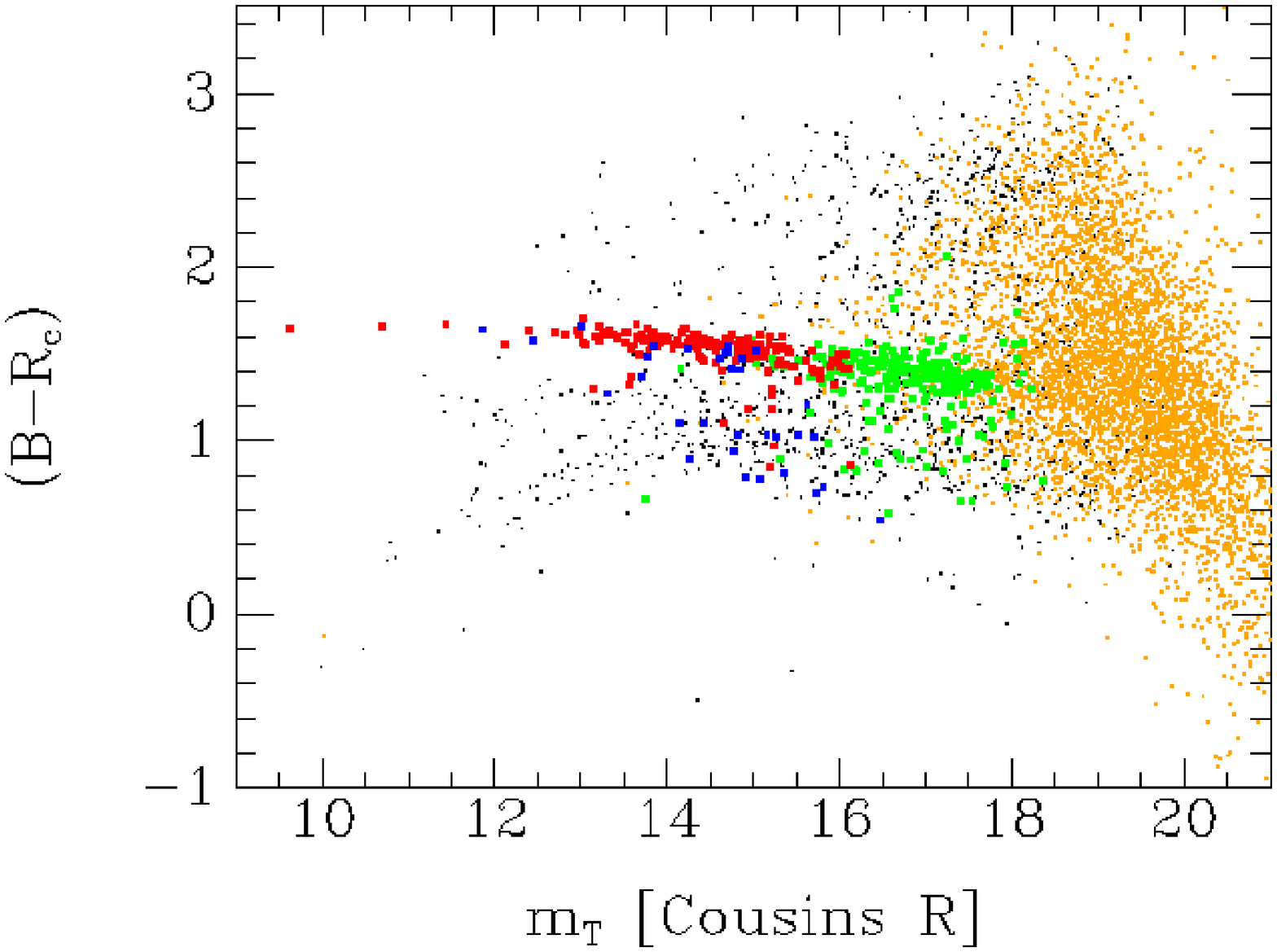,width=14truecm,angle=-90,clip=}}
{ Color-magnitude relation for the central 75 arcmin $\times$ 80 arcmin
of the Coma cluster. Both stars and galaxies are included on this
figure, a total of 15370 objects. The lack of objects in the upper
right hand corner is due to the combination of the magnitude limits 
in B and $R_{\rm c}$. Small black points -- stars;
black [blue] boxes -- spirals and irregulars, confirmed members;
dark grey [red] boxes -- E and S0 galaxies, confirmed members;
light grey [green] boxes -- unclassified confirmed members;
small grey [orange] points -- other galaxies in the field, members and
non-members. The photometric data are from J\o rgensen (2000).
The redshift data are from J\o rgensen \& Hill (2000).
}

Color-color relations in the optical are in Figure~15 shown for the
same sample of objects as shown in Figure~14. The confirmed members
of the cluster form a tight relation in these two color-color relations.
This is in agreement with predictions from stellar population models
(see Section 4), which predicts the optical colors to be degenerate
in age and metallicity. 
This is seen on Figure 15, left panel, where
the same models as shown on Figure 13 are overplotted.
The lines of constant age fall right on top of
the lines of constant metallicity, forming a single line along
the ridge of the location of the E and S0 galaxies.
This means that the optical colors alone cannot
be used to derive the mean ages and the mean metallicities.
A younger age will lead to bluer colors, but a similar change in 
colors can be caused by a lower metallicity. 

\midinsert
\vbox{%
\centerline{\psfig{figure=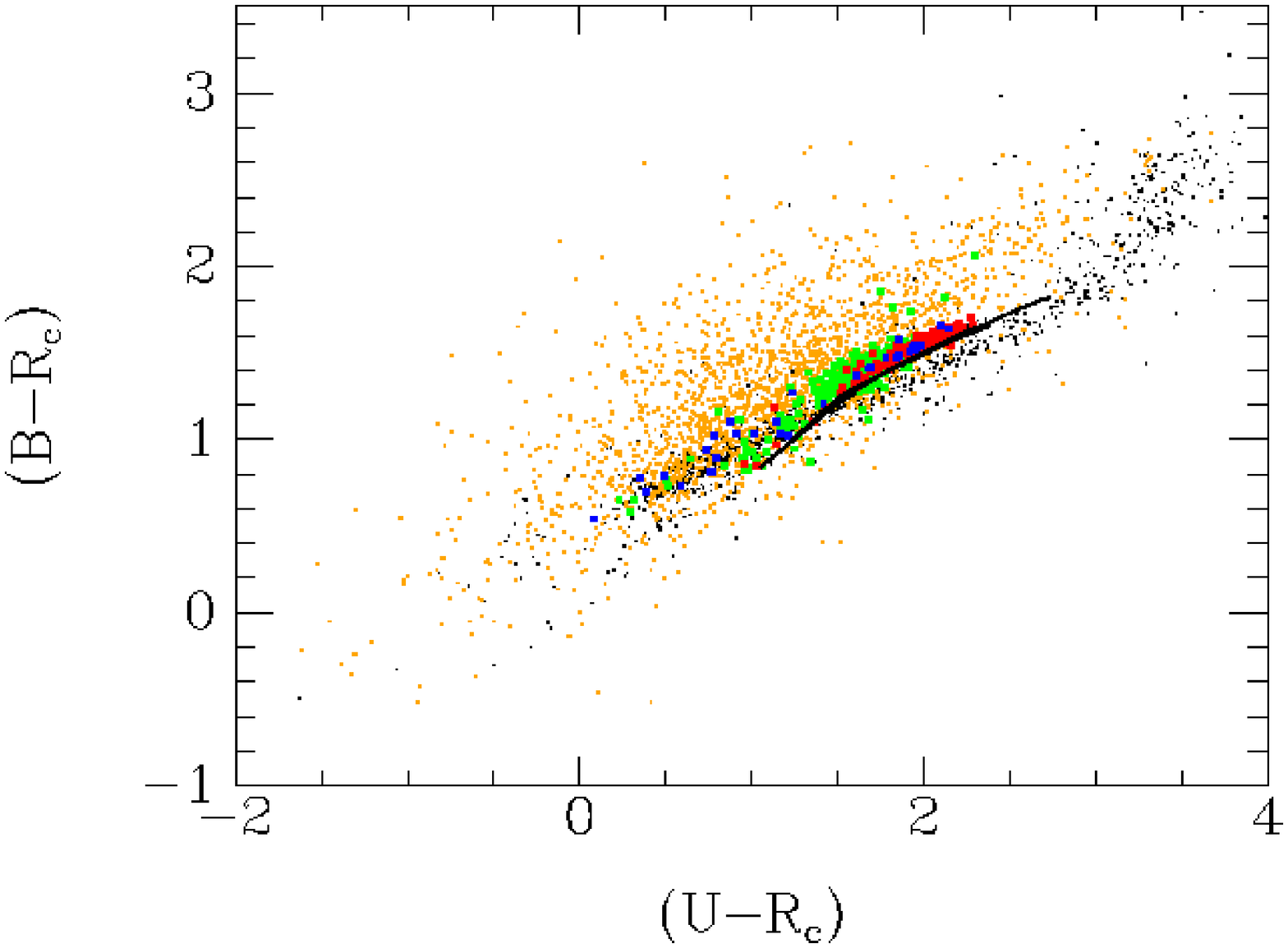,width=6.5truecm,angle=-90,clip=}
\psfig{figure=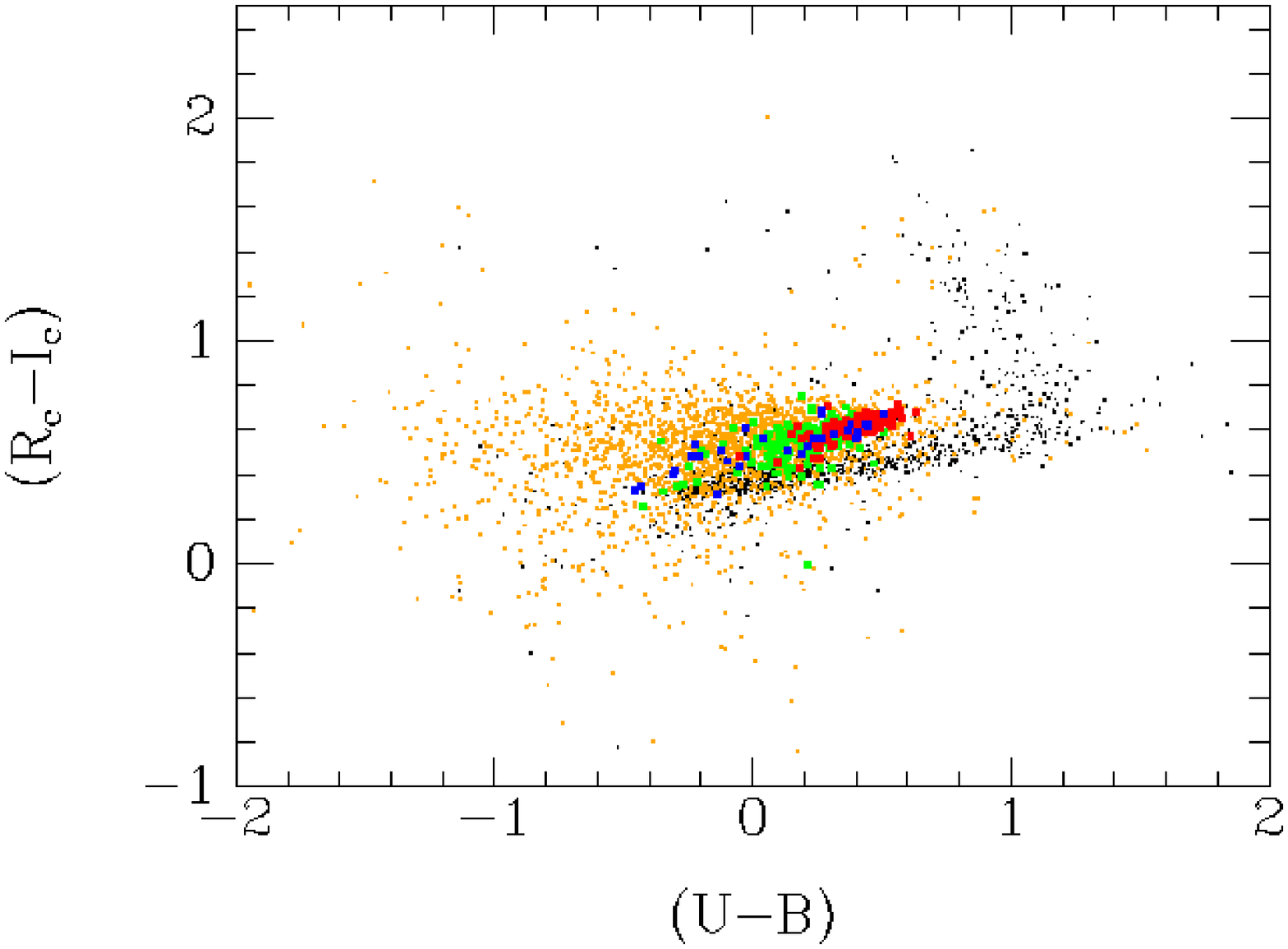,width=6.5truecm,angle=-90,clip=}}
\vskip2truemm
\bcaption{15}
{Optical color-color relations for the Coma cluster.
Symbols as in Figure~14. Solid line on the left panel -- SPP models from
Vazdekis et al.\ (1996), see text.
}}
\endinsert

While the optical color-color diagrams are degenerate in age and
metallicity, the combination of one optical color and 
one optical-infrared color may be used to break the degeneracy.
An example of this is shown in Figure~16 for a small sample of 
E and S0 galaxies in the Coma cluster.
The optical color (U$-$B) and the optical-infrared color (V$-$K)
are both sensitive to both age differences and metallicity differences.
However, (U$-$B) is more sensitive to the age differences than to
metallicity differences, while the opposite is the case for (V$-$K).
The stellar population models in the near infrared (near-IR), JHK, 
in this case 
the K-band, are still rather uncertain, but the technique is promising 
for studies of faint high redshift galaxies for which spectroscopy
would come at a very high cost of telescope time at 8-meter class
telescopes. 
The apparently rather low mean age of the Coma cluster galaxies
as estimated from Figure~16 is in fact in agreement with
results based on spectroscopic information (see J\o rgensen 1999).

\WFigure{16}{\psfig{figure=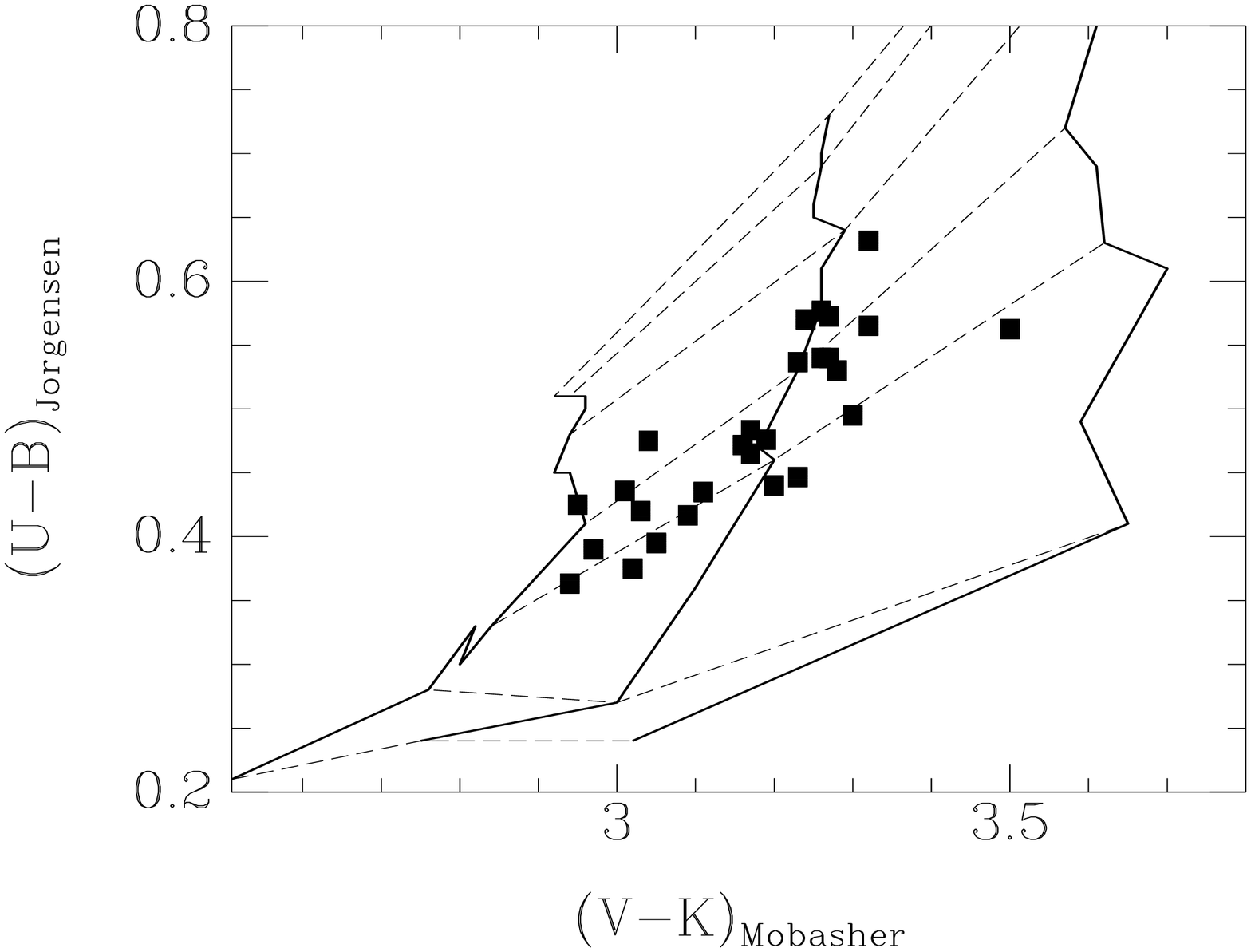,width=10truecm,angle=-90,clip=}}
{Optical-infrared color-color relation for the Coma cluster.
The data are from J\o rgensen (2000) and Mobasher et al.\ (1999).
The lines represent stellar population models by Vazdekis et al.\ 
(1996). Solid lines -- metallicities of [Fe/H]=$-$0.4, 0.0 and 0.4;
the lowest metallicity leads to the smallest (V$-$K).
Dashed lines -- ages of 1, 2, 5, 8, 12, 15, and 17 Gyr; the largest
ages lead to the largest (U$-$B). }

\vskip2truemm

\section{6. THE FUNDAMENTAL PLANE}

The Fundamental Plane (FP) is a relation that combines
surface photometry with spectroscopy.
We will discuss this relation both at low and at high redshift.

The FP (Djorgovski \& Davis 1987; Dressler et al.\ 1987)
is the relation
$$\log\re = \alpha\log\sigma + \beta\log\Ie + \gamma \enspace , \eqno(12)$$
where $\sigma$ is the line-of-sight stellar velocity dispersion
for the galaxy in question.
In other words, the measured values of $\log\re$, $\log\Ie$ 
and $\log\sigma$ for a sample of E and S0 galaxies
do not populate this 3-parameter space evenly, but are limited
to a thin plane.

The velocity dispersion $\sigma$ is determined from spectroscopy,
see Figure~17.
The absorption lines in galaxies are broadened due to the internal
motions of the stars in the galaxy.
To determine how much the stars are moving, it is necessary to know what the
spectrum of the galaxy would be if all the stars were at rest
with respect to each other.
This is approximated by the spectrum of af K giant star.
This template star spectrum is broadened by a Gaussian broadening function
until it matches the galaxy spectrum.
The velocity dispersion of the galaxy is then the dispersion $\sigma$
of the broadening function.
In more precise terms, this determination of $\sigma$ can be done using the
Fourier fitting method (Franx, Illingworth \& Heckman 1989a) or the
Fourier quotient method (Sargent et al.\ 1977).

\WFigure{17}{%
\psfig{file=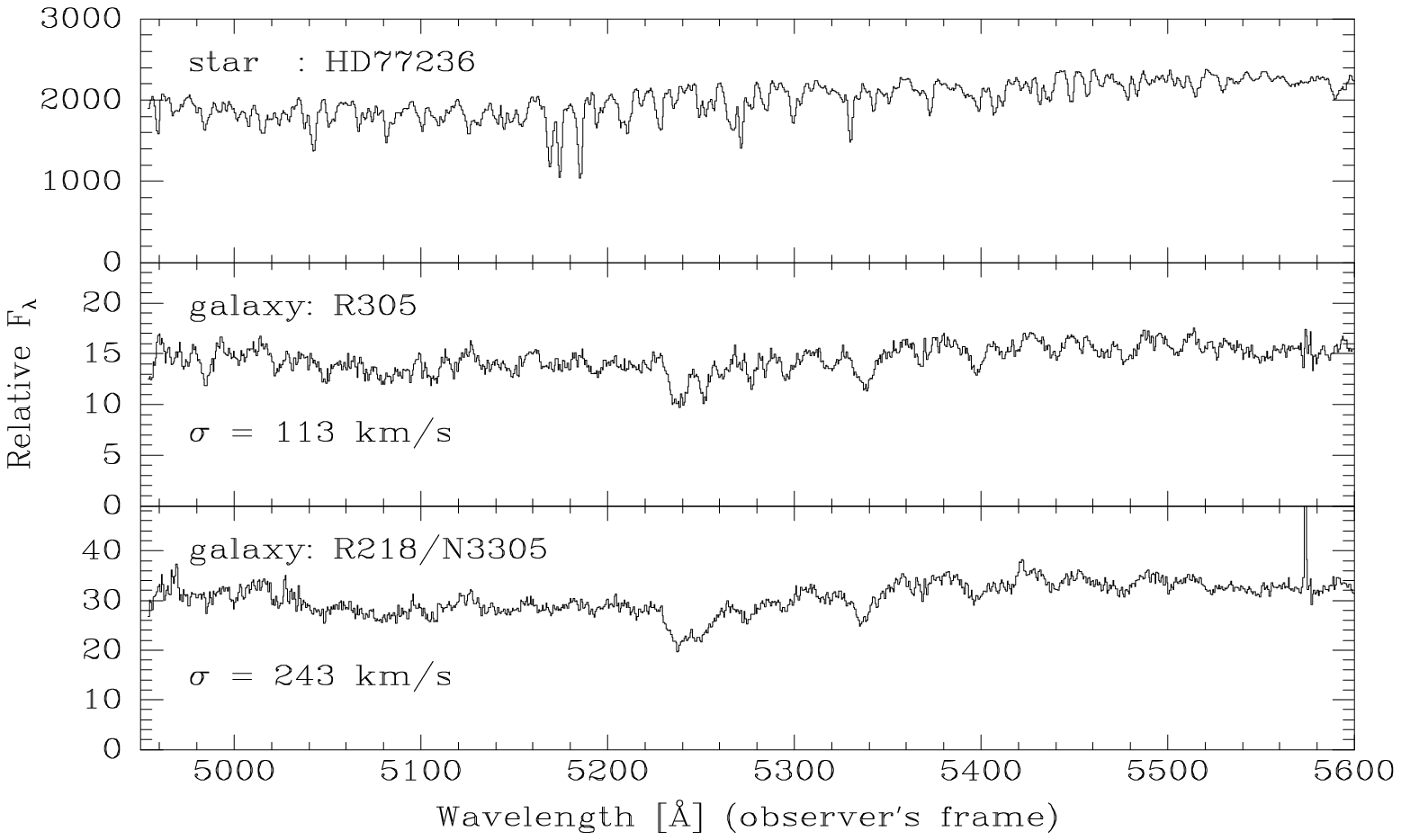,width=\dimen9,clip=}
}
{Illustration of the effect of the velocity dispersion $\sigma$.
The top panel shows a K giant star in our own Galaxy.
This star is representative of the stellar populations in E or S0 galaxies.
The two lower panels show E or S0 galaxies in the HydraI cluster.
The ${\rm Mg}\,{\rm b}$ absorption line triplet at 5177 {\AA} (rest frame)
is broadened in the galaxy spectra. The instrumental resolution is 79 km/s.}

\subsection{6.1 The interpretation of the Fundamental Plane}

The physics behind the FP can be illuminated by some simple arguments
(Djorgovski, de Carvalho \& Han 1988; Faber et al.\ 1987).
Consider the virial theorem for the stars in the galaxy
$$
{GM \over \Rangle} = 2 \, {\Vanglesq \over 2} \enspace , \eqno(13)
$$
We relate the observable quantities $\re$, $\sigma$ and $\Ie$
to the `physical' quantities $\Rangle$, $\Vanglesq$ and luminosity $L$ through
$$
\re      = \kR \Rangle      \enspace , \quad\quad
\sigma^2 = \kV \Vanglesq    \enspace , \quad\quad
L        = \kL \Ie^{} \re^2 \enspace , \eqno(14)
$$
The parameters $\kR$, $\kV$, and $\kL$ reflect the
density structure, kinematical structure, and luminosity structure
of the given galaxy.
If these parameters are constant, the galaxies
constitute a {\it homologous\/} family.
Homology means that structure of small and big galaxies is the same.

Combining Equation (13) and (14) gives
$$
\re = \kS (M/L)^{-1} \sigma^2 \Ie^{-1} \enspace , \quad\quad
\kS = (G\kR\kV\kL)^{-1} \enspace . \eqno(15)
$$
For homology $\kS$ will be constant.
When this relation is compared to the observed FP,
$$
\re = {\rm constant} \cdot \sigma^{1.24\pm0.07} \Ie^{-0.82\pm0.02}
\eqno(16)
$$
(J{\o}rgensen, Franx \& Kj{\ae}rgaard 1996, in Gunn $r$),
it is seen that the coefficients of the FP
are not 2 and $-1$ as expected from homology and constant mass-to-light ratios.
The product $\kS (M/L)^{-1}$ cannot be constant,
but has to be a function of $\sigma$ and $\Ie$.
A non-constant $\kS (M/L)^{-1}$ can be explained by
a systematic deviation from homo\-logy ($\kS$ varies), or
a systematic variation of the $M/L$ ratios, or both.
When homology is assumed, the observed FP coefficients give the relation
$$
M/L_{\rm r} \propto M^{0.24\pm0.03} \enspace , \eqno(17)
$$
(J\o rgensen et al.\ 1996).
The interpretation of the FP is still a matter of debate.
The $M/L \propto M^b$ interpretation seems to be the most favored one,
although there is some evidence that non-homology may play a role too
(e.g.\ Hjorth \& Madsen 1995; Pahre, de Carvalho \& Djorgovski 1998).

\subsection{6.2 The evolution of the Fundamental Plane as a function of
redshift}

The Fundamental Plane can be used to study the evolution of
galaxies as a function of redshift. As explained above,
the FP may be interpreted
as a relation between the masses and the $M/L$ ratios of the galaxies.
Under the assumption that the masses do not change with
redshift, e.g.\ no merging takes place, the evolution of the FP zero 
point with redshift can be interpreted as the evolution of the $M/L$ 
ratios. 

Several authors have studied the FP for clusters at redshifts higher
than 0.1, see Table~5 for clusters and references.
Additional studies by Pahre, Djorgovski \& de Carvalho (1999) and 
Kelson et al.\ (1999) are soon to be published in refereed journals.

\midinsert
$$\vbox{\tabfont
\halign{
\huad #\hfil&           
\hfil\huad #&           
\hfil\huad #&           
\huad #\hfil\cr         
\multispan{4}{{\tbold Table 5.}~~Fundamental Plane studies of cluster with $z>0.1$\hfil}\cr
\tablerule
Cluster & $z$ & $N_{\rm gal}$ & Reference  \cr
\tablerule
A2218       & 0.18 & 9 & J\o rgensen \& Hjorth (1997), \cr
            & & & J\o rgensen et al.\ (1999) \cr
A665        & 0.18 & 6 & J\o rgensen \& Hjorth (1997), \cr
            & & & J\o rgensen et al.\ (1999) \cr
CL1358+62   & 0.33 & 10 & Kelson et al.\ (1997) \cr
MS1512+36   & 0.37 & 2 & Bender et al.\ (1998) \cr
A370        & 0.37 & 7 & Bender et al.\ (1998) \cr
CL0024+16   & 0.39 & 8 & van Dokkum \& Franx (1996) \cr
MS2053$-$04 & 0.58 & 5 & Kelson et al.\ (1997) \cr
MS1054$-$03 & 0.83 & 8 & van Dokkum et al.\ (1998) \cr
\tablerule
}}
$$
\endinsert

As examples of high redshift studies of the FP we show in Figure~18
the FP for the Coma cluster and for five clusters with redshift larger
than 0.1. The data for the Coma are from J\o rgensen (1999) and
J{\o}rgensen et al.\ (1995). 
Abell 2218 and Abell 665 are discussed
by J\o rgensen et al.\ (1999). The sources for the rest of the
clusters are given on the figure.

\WFigure{18}{\psfig{figure=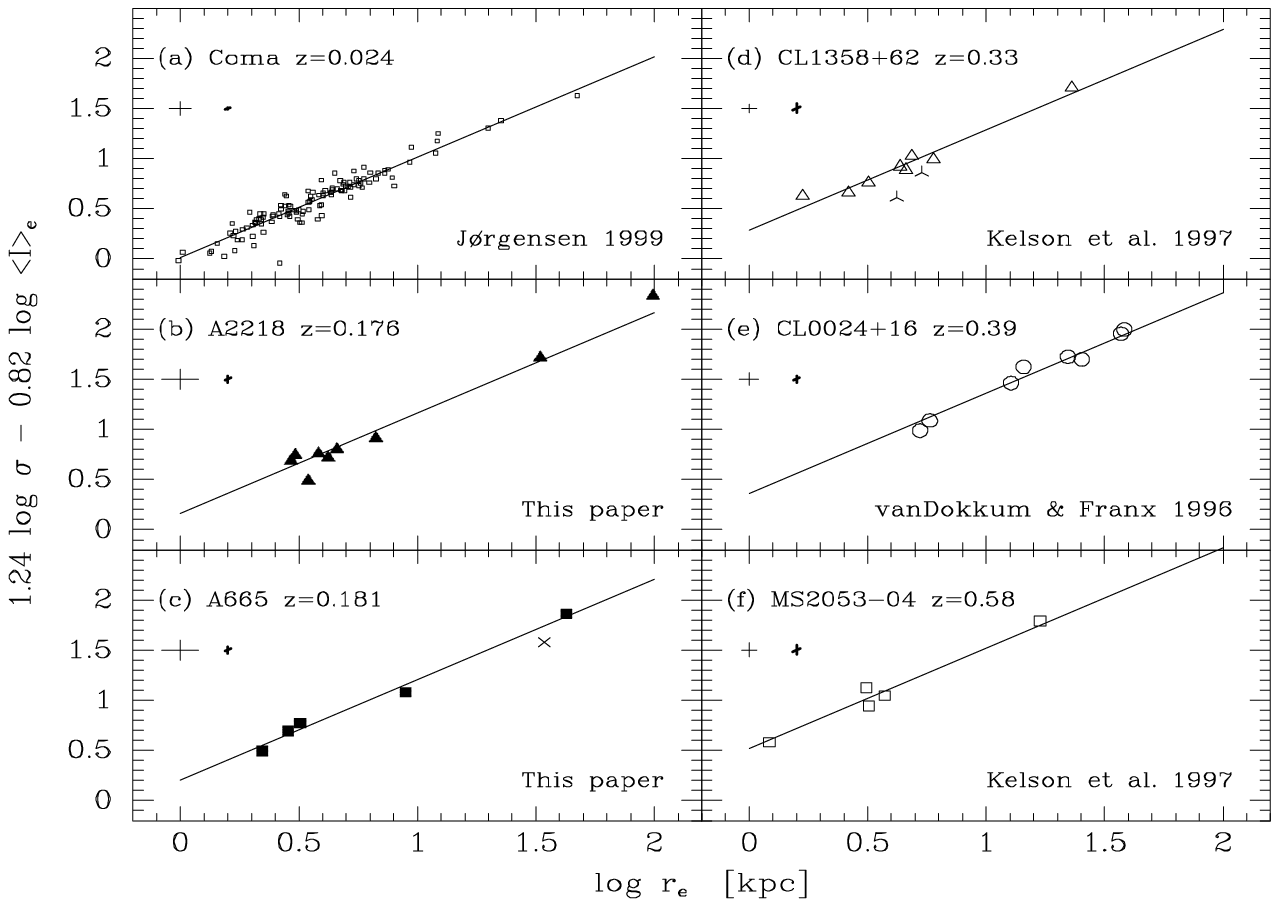,width=13truecm,angle=0,clip=}}
{ The FP edge-on for Coma, A2218, A665, CL1358+62, CL0024+16,
and  MS2053$-$04.
The sources of the data are given on the panels
(`This paper' refers to J\o rgensen et al.\ 1999).
The skeletal symbols on panel (c) and (d) are the E+A galaxies.
The photometry is calibrated to Gunn $r$ in the rest frames of
the clusters. 
The mean surface brightness $\log \Ie = -0.4(\mue -26.4)$ 
is in units of $\rm L_{\odot} / pc^2$ (cf.\ Section 2.4.1).
The photometry is not corrected for the dimming
due to the expansion of the Universe. The effective radii are
in kpc ($\Ho50$ and $q_{\rm o} = 0.5$).
The solid lines are the FPs with coefficients adopted from 
J{\o}rgensen et al.\ (1996), 
and with zero points derived from the data presented in the figure.
Typical error bars are given on the panels; the thin and thick error 
bars show the random and systematic uncertainties, respectively.
(From J\o rgensen et al.\ 1999.)
}

 From the change in the zero point of the FP as a function of redshift
J\o rgensen et al.\ (1999),
in agreement with other studies, find that the $M/L$ ratios of the
E and S0 galaxies change very slowly with redshift. The sample of 
clusters shown on Figure~18 span about half of the current age
of the Universe (for $q_0=0.5$).
Under the assumption that the galaxies evolve passively over this
time interval, e.g.\ no merging and no formation of new E and S0 
galaxies, then
it is possible to put limits of the redshift at which the majority
of the stars were formed. This redshift is called the formation
redshift. The study by J\o rgensen et al.\ (1999) as well as other
studies conclude that the formation redshift is larger than about
2.5 for $q_0=0.5$ and larger than about 1.5 for $q_0 =0.15$.

It is important to keep in mind that the assumption regarding passive 
evolution represents a very simplified view of the galaxy evolution.
Most likely the real evolution since $z \approx 0.6$ cannot be modeled 
with passive evolution.
If the galaxies experience on-going star formation, then 
the observed evolution will appear smaller than for passive evolution
because the galaxies are continuously forming young bright stars.
A similar effect can be caused by a series of smaller bursts of
star formation.
Finally, the interactions, the possible merging and the morphological
evolution of the galaxies over the last half of the age of the 
Universe cannot be ignored.
Some of the E and S0 galaxies that we observe in low redshift clusters 
may not have ended up in the samples if we could have observed those 
clusters at a much earlier stage in their evolution, 
simply because some of the E 
and S0 galaxies may have formed recently by merging of spiral galaxies.

\vskip2truemm

\section{7. SUGGESTED FUTURE PROJECTS}

We end this paper by a brief summary of some of the projects
that may be carried out building on the techniques and results
discussed in this paper. We concentrate on projects that involve
photometry only. Some of the projects may be carried out
using existing archive data from HST.

\subsection{Evolution of morphology as a function of redshift}

Dressler et al.\ (1997) have recently used HST/WFPC2 data of 
clusters to study the morphology-density relation as a function of 
redshift. Dressler et al.\ found that the fraction of S0 galaxies
is lower at high redshift than at low redshift, while the fraction
of spirals is higher at high redshift than at low redshift.
However, this study is based on the traditional method of classifying 
galaxies.
We suggest that a quantitative approach is taken to any study of
the morphology. Of special interest would be to study
how the relative disk luminosities $\LD /\Ltot$ for E 
and S0 galaxies evolve with redshift, both in clusters and in
the field. The first study of this kind was done for the cluster
CL0024+16 ($z\!=\!0.39$) by Bergmann \& J\o rgensen (1999) who found that the
$\LD /\Ltot$ distribution for the E and S0 galaxies in CL0024+16
shows a paucity of disk-dominated galaxies when compared to
the Coma cluster. 

It would be valuable to apply the same technique and derive
$\LD /\Ltot$ for a larger sample of cluster and field galaxies
at redshifts larger than 0.1 to establish the possible evolution
of the distribution of $\LD /\Ltot$.
This may be done using the HST/WFPC2 archive data.

\subsection{Studies of global colors}

There have been many studies of the global colors of galaxies
as a function of redshift (e.g.\ Stanford, Eisenhardt \& Dickinson 1998;
Bower, Kodama \& Terlevich 1998; Kodama et al.\ 1998 and references in these
papers).
However, most studies concentrate on the optical colors
of the galaxies, while the near-IR (JHK) data is very sparse.
The study by Stanford et al.\ includes near-IR data and addresses
the question of how the color-magnitude relations for the
near-IR colors evolve with redshift.

The combination of optical and optical-infrared colors may be used to 
break the age-metallicity degeneracy (see Section 5). 
Observations of low redshift E and S0 galaxies in the near-IR
may be used to establish the zero redshift properties and the 
methods and models needed to break the age-metallicity degeneracy.
For high redshift galaxies ($z>0.5$), the near-IR photometry
may be obtained with 8-meter class telescopes with superior spatial 
resolution.
Such data will give the possibility of studying the mean ages
and mean metallicities as functions of redshift for significantly
fainter galaxies than it is currently possible by obtaining spectroscopy
with 8-meter class telescopes.

\vfill\eject

\subsection{Color gradients}

While we have not discussed color gradients in this paper, color
gradients provide an alternative method of studying galaxy 
evolution. The color gradients in E and S0 galaxies reflect
underlying radial gradients in the metallicity (and maybe the age) of 
the stellar populations.
Models for galaxy formation predict the sizes of these gradients.
In general, the predicted gradients are steeper for models based on
a monolithic collapse (Carlberg 1984) than for models based on 
the merger hypothesis (White 1980).
Determination of color gradients for high redshift galaxies requires
high signal-to-noise data with very good spatial resolution. 
Several of the rich galaxy clusters observed with HST/WFPC2 have
sufficiently high signal-to-noise data that a study may be carried
out using the available archive data.

\vskip4truemm

ACKNOWLEDGEMENTS.
It is a pleasure to thank the organizers for a successful and stimulating
school.
Support from the Nordic Research Academy (REF 99.10.003-B) for the course is
kindly acknowledged,
as well as support from
Nato Scientific and Environmental affairs division linkage
grant, Computer network supplement 97 46622 Re CRG.LG 972172
for the Internet connection.

The data used in this paper were obtained at
the Nordic Optical Telescope, the Danish 1.5-meter Telescope 
LaSilla, the Kitt Peak National Observatory 4-meter Telescope,
the Multi-Mirror Telescope, the McDonald Observatory 0.8-meter
and 2.7-meter Telescopes, and the Hubble Space Telescope.
We thank the telescope allocation committees for granting time
to these project, and the staff at NOT, ESO/LaSilla, KPNO, MMT and
McDonald Observatory for assistance during the observations.

\References
\ref
 Abraham R. G., Valdes F., Yee H. K. C., van den Bergh S., 1994, ApJ, 432, 75
\ref
 Abraham R. G., van den Bergh S., Glazebrook K., Ellis R. S., Santiago B. X.,
 Surma P., Griffiths R. E., 1996, ApJS, 107, 1
\ref
 Bender R., M{\"o}llenhoff C., 1987, A\&A, 177, 71
\ref
 Bender R., D{\"o}bereiner S., M{\"o}llenhoff C., 1988, A\&AS, 74, 385
\ref 
 Bender R., Saglia R. P., Ziegler B., Belloni R.,
 Greggio L., Hopp U., Bruzual G., 1998, ApJ, 493, 529
\ref
 Bender R., Surma P., D{\"o}bereiner S., M{\"o}llenhoff C., Madejsky R., 1989,
 A\&A, 217, 35
\ref
 Bergmann M., J\o rgensen I., 1999,
 in Galaxy Dynamics,
 ASP Conference Series Vol.\ 182,
 eds.\ D. R. Merritt, M. Valluri, J. A. Sellwood,
 p.\ 505
\ref
 Bertelli G., Bressan A., Chiosi C., Fagotto F., Nasi E., 1994, A\&AS, 106, 275
\ref 
 Bower R. G., Kodama T., Terlevich A., 1998, MNRAS, 299, 1193
\ref
 Busko I., 1996,
 in Astronomical Data Analysis Software and Systems V,
 ASP Conference Series Vol.\ 101,
 eds.\ G. H. Jacoby, J. Barnes,
 p.\ 139
\ref 
 Carlberg R. G., 1984, ApJ, 286, 403
\ref
 Carter D. 1987, ApJ, 312, 514
\ref
 Cawson M. C., 1983, PhD Thesis, University of Cambridge
\ref
 Davis L. E., Cawson M., Davies R. L., Illingworth G., 1985, AJ, 90, 169
\ref
 de Vaucouleurs G., 1948, Ann.\ d'Ap., 11, 247
\ref
 Djorgovski S., Davis M., 1987, ApJ, 313, 59
\ref
 Djorgovski S., de Carvalho R., Han M.-S., 1988,
 in The Extragalactic Distance Scale,
 ASP Conference Series Vol.\ 4,
 eds.\ van den Bergh S., Prichet C. J.,
 p.\ 329
\ref
 Dressler A., Lynden-Bell D., Burstein D., Davies R. L., Faber S. M.,
 Terlevich R. J., Wegner G., 1987, ApJ, 313, 42
\ref 
 Dressler A., Oemler A. Jr., Couch W. J., Smail I.,
 Ellis R. S., Barger A., Butcher H., Poggianti B. M.,
 Sharples R. M., 1997, ApJ, 490, 577
\ref
 Faber S. M., Dressler A., Davies R. L., Burstein D.,
 Lynden-Bell D., Terlevich R. J., Wegner G., 1987,
 in Nearly Normal Galaxies,
 ed.\ Faber S. M.
 Springer, New York, p.\ 175
\ref
 Faber S. M., Tremaine S., Ajhar E. A., Byun Y.-I., Dressler A., 
 Gebhardt K., Grillmair C., Kormendy J., Lauer T. R., Richstone D., 1997, 
 AJ, 114, 177
\ref 
 Franx M., Illingworth G., Heckman T., 1989a, ApJ, 344, 613
\ref 
 Franx M., Illingworth G., Heckman T., 1989b, AJ, 98, 538
\ref
 Hjorth J., Madsen J., 1995, ApJ, 445, 55
\ref
 Jedrzejewski R. I., 1987, MNRAS, 226, 747
\ref
 J\o rgensen I., 1999, MNRAS, 306, 607
\ref
 J\o rgensen I., 2000, in preparation
\ref
 J\o rgensen I., Franx M., 1994, ApJ, 433, 553 (JF94)
\ref
 J\o rgensen I., Hill G., 2000, in preparation
\ref
 J\o rgensen I., Hjorth J., 1997,
 in Galaxy Scaling Relations: Origins, Evolution and Applications,
 eds.\ L. N. da Costa, A. Renzini, Springer-Verlag,
 p.\ 175
\ref
 J\o rgensen I., Franx M., Hjorth J., van Dokkum P., 1999, MNRAS,
 308, 833
\ref
 J\o rgensen I.,  Franx M., Kj{\ae}rgaard P.,  1992,
 A\&AS, 95, 489
\ref
 J\o rgensen I.,  Franx M., Kj{\ae}rgaard P.,  1995, MNRAS, 273, 1097
\ref
 J\o rgensen I.,  Franx M., Kj{\ae}rgaard P.,  1996, MNRAS, 280, 167 
\ref
 Kauffmann G., Charlot S., 1998, MNRAS, 294, 705
\ref
 Kelson D. D., van Dokkum P. G., Franx M., Illingworth G. D., 
 Fabricant D., 1997, ApJ, 478, L13
\ref
 Kelson D. D., Illingworth G. D., Franx M., van Dokkum P. G., 1999,
 in The High Redshift Universe: Galaxy Formation and Evolution at High Redshift,
 ASP Conference Series, in press
\ref
 Kodama T., Arimoto N., Barger A. J., Arag\'{o}n-Salamanca A.,
 1998, A\&A, 334, 99
\ref
 Kormendy J., Djorgovski S., 1989, ARA\&A, 27, 235
\ref
 Milvang-Jensen B., J{\o}rgensen I., 2000, in preparation
\ref
 Mobasher B., Guzm\'{a}n R., Arag\'{o}n-Salamanca A., Zepf S., 1999,
 MNRAS, 304, 225
\ref
 Naim A., Lahav O., Sodre L. Jr., Storrie-Lombardi M. C.,
 1995, MNRAS, 275, 567
\ref
 Nieto J.-L., Bender R., 1989, A\&A, 215, 266
\ref
 Okamura S., 1988, PASP, 100, 524
\ref
 Pahre M. A., de Carvalho R. R., Djorgovski S. G., 1998, AJ, 116, 1606
\ref
 Pahre M. A., Djorgovski S. G., de Carvalho R. R., 1999,
 in Star Formation in Early Type Galaxies,
 ASP Conference Series Vol.\ 163,
 eds.\ P. Carral, J. Cepa,
 p.\ 17
\ref
 Peletier R. F., Davies R. L., Illingworth G. D., Davis L. E., Cawson M., 1990,
 AJ, 100, 1091 
\ref
 Press W. H., Teukolsky S. A., Vetterling W. T., Flannery B. P.,
 1992, Numerical Recipes, Cambridge University Press, New York
\ref
 Richter O.-G., 1989, A\&AS, 77, 237
\ref
 Rix H.-W., White S. D. M., 1990, ApJ, 362, 52
\ref
 Saglia R. P., Bertschinger E., Baggley G., Burstein D., Colless M.,
 Davies R. L., McMahan Jr. R. K., Wegner G., 1993,
 MNRAS, 264, 961
\ref
 Salpeter E. E., 1955, ApJ, 121, 161
\ref
 Sargent W. L. W., Schechter P. L., Boksenberg A., Shortridge K., 1977,
 ApJ, 212, 326
\ref
 Stanford S. A., Eisenhardt P. R., Dickinson M., 1998, ApJ, 492, 461
\ref
 Thuan T. X., Gunn J. E., 1976, PASP, 88, 543
\ref
 van Dokkum P. G., Franx M., 1996, MNRAS, 281, 985
\ref
 van Dokkum P. G., Franx M., Kelson D.\ D., 
 Illingworth G. D., 1998, ApJ, 504, L17
\ref
 Vazdekis A., Casuso E., Peletier R. F., Beckman J. E.,
 1996, ApJS, 106, 307
\ref
 White S. D. M., 1980, MNRAS, 191, 1p
\ref
 Worthey G., 1994, ApJS, 95, 107
\ref
 Worthey G., Trager S. C., Faber S., 1995,
 in Fresh Views on Elliptical Galaxies,
 ASP Conference Series Vol.\ 86,
 eds.\ A. Buzzoni, A. Renzini, A. Serrano,
 p.\ 203

\bye